\documentclass[preprint]{elsarticle}

\usepackage[utf8]{inputenc}
\usepackage{url}
\usepackage{amsmath}
\usepackage{amssymb}
\usepackage{bbm}
\usepackage{xspace}
\usepackage{siunitx}
\usepackage{todonotes}
\usepackage{algorithm}
\usepackage{algpseudocode}
\usepackage[toc,page]{appendix}
\usepackage{amsthm}
\usepackage{pdflscape}
\usepackage{subcaption}
\usepackage[section]{placeins}
\usepackage{lineno}
\usepackage[colorlinks=true,allcolors=blue]{hyperref}
\usepackage[capitalize]{cleveref}
\usepackage{booktabs}
\usepackage{amsthm}

\theoremstyle{plain}
\newtheorem*{remark}{Remark}

\theoremstyle{remark}

\numberwithin{equation}{section}

\graphicspath{{./img/}}

\newcommand{\dumux}{DuMu\textsuperscript{x}\xspace}

\renewcommand{\vec}[1]{\boldsymbol{#1}}

\renewcommand{\div}{\nabla\cdot\!}
\newcommand{\grad}{\nabla\!}

\newcommand{\meas}[1]{\lvert{#1}\rvert}
\newcommand{\norm}[1]{\|#1\|}

\newcommand{\up}[1]{\mathrm{#1}}
\newcommand{\sthreed}{\up{3D}}
\newcommand{\soned}{\up{1D}}

\newcommand{\xOne}{x_1}
\newcommand{\xTwo}{x_2}
\newcommand{\xThree}{x_3}

\newcommand*\patchAmsMathEnvironmentForLineno[1]{%
	\expandafter\let\csname old#1\expandafter\endcsname\csname #1\endcsname
	\expandafter\let\csname oldend#1\expandafter\endcsname\csname end#1\endcsname
	\renewenvironment{#1}%
	{\linenomath\csname old#1\endcsname}%
	{\csname oldend#1\endcsname\endlinenomath}}%
\newcommand*\patchBothAmsMathEnvironmentsForLineno[1]{%
	\patchAmsMathEnvironmentForLineno{#1}%
	\patchAmsMathEnvironmentForLineno{#1*}}%
\AtBeginDocument{%
	\patchBothAmsMathEnvironmentsForLineno{equation}%
	\patchBothAmsMathEnvironmentsForLineno{align}%
	\patchBothAmsMathEnvironmentsForLineno{flalign}%
	\patchBothAmsMathEnvironmentsForLineno{alignat}%
	\patchBothAmsMathEnvironmentsForLineno{gather}%
	\patchBothAmsMathEnvironmentsForLineno{multline}%
}

\renewcommand{\citet}{\citep}

\algdef{SE}[VARIABLES]{Variables}{EndVariables}
   {\algorithmicvariables}
   {\algorithmicend\ \algorithmicvariables}
\algnewcommand{\algorithmicvariables}{\textbf{variables}}
\algrenewcommand\textproc{\texttt}

\begin{document}

\title{Projection-based resolved interface mixed-dimension
method for embedded tubular network systems}
\author[1,2]{Timo Koch}
\ead{timokoch@uio.no}

\address[1]{Department of Mathematics, University of Oslo, Norway}
\address[2]{Department of Hydromechanics and Modelling of Hydrosystems, University of Stuttgart, Germany}

\begin{abstract}
We present a flexible discretization technique for computational models of thin tubular networks
embedded in a bulk domain, for example a porous medium.
These systems occur in the simulation of fluid flow in vascularized biological tissue,
root water and nutrient uptake in soil, hydrological or petroleum wells in rock formations,
or heat transport in micro-cooling devices. The key processes, such as heat and mass transfer, are usually
dominated by the exchange between the network system and the embedding domain. By explicitly resolving
the interface between these domains with the computational mesh, we can accurately describe these processes.
The network is efficiently described by a network of line segments. Coupling terms are evaluated by projection
of the interface variables. The new method is naturally applicable for nonlinear and time-dependent problems
and can therefore be used as a reference method in the development of novel implicit interface 1D-3D methods
and in the design of verification benchmarks for embedded tubular network methods. Implicit interface,
not resolving the bulk-network interface explicitly have proven
to be very efficient but have only been mathematically analyzed for linear elliptic problems so far.
Using two application scenarios, fluid perfusion of vascularized tissue and root water uptake from soil,
we investigate the effect of some common modeling assumptions
of implicit interface methods numerically.
\end{abstract}


\maketitle

\section{Introduction}
\label{sec:intro}

\begin{figure}[htb!]
  \includegraphics[width=1.0\textwidth]{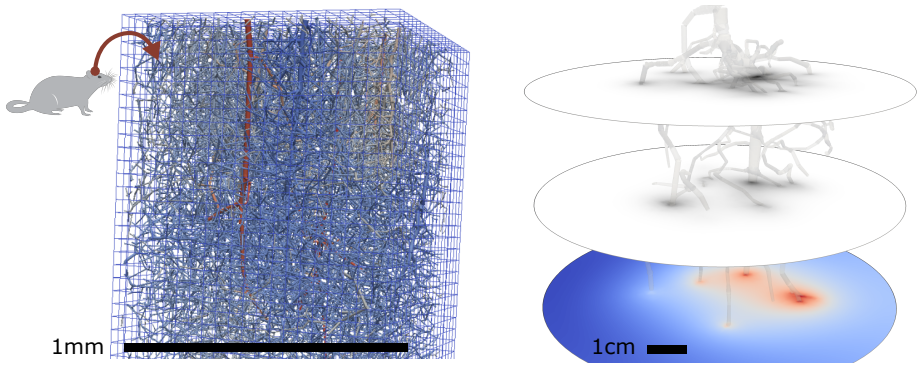}%
  \caption{\textbf{Examples of systems with embedded network structures.} Left, blood vessel network simulation (geometry and boundary conditions from~\citep{Reichold2009})
           in the rat brain cortex coupled with the embedding tissue. Color signifies simulated blood pressure.
           A grey matter tissue volume of \SI{1}{\cubic\milli\m} may contain around 10'000 blood vessels.
           Right, simulation of root water and tracer uptake from soil by a young lupine root system,
           image adapted from~\citep{Koch2019cdumux} (license: CC BY 4.0).}
  \label{fig:introduction}
\end{figure}

There is a strong demand for efficient and accurate models describing flow and transport
processes in porous media with embedded tubular network systems, such as vascularized biological tissue,
plant root system growing in soil, hydrological, geothermal or petroleum wells in rock formations.
Reduced models are necessary due to the computational complexity arising from the large number
of network segments (for example about 10'000 blood vessels in a \SI{1}{\cubic\milli\m}
cube of gray matter brain tissue~\citep{Blinder2013}, or hundreds of meters of cumulative root length
in a $60$-day-old maize root system~\citep{Leitner2014b}) and the small diameter of the tubes with respect to
the entire computational domain
(for example wells of \SI{10}{\centi\m} diameter in a \si{\kilo\m}-scale reservoir).
Two motivational examples of mixed-dimensional simulations are shown in \cref{fig:introduction}.

Various methods have been developed recently to numerically solve coupled mixed-dimensional
partial differential equations (PDEs) that arise from flow and transport models in such systems.
Typically, flow and transport in the embedded tubular network system are described by one-dimensional equations
posed on a network of (center-)line segments. These networks are embedded into the surrounding bulk medium, often
porous media, which are described by three-dimensional equations. Network and bulk PDEs are coupled by
source terms that depend on state variables from both domains. The different numerical techniques
differ in the way they deal with the dimensional gap of $2$ between network and bulk domain.
The source term contribution in the embedding bulk medium can
be described by line source terms~\citep{Hsu1989green,dangelo2008,DAngelo2012,cattaneo2014computational,Gjerde2018},
surface source terms~\citep{koeppl2018,Laurino2019} or volume source terms~\citep{Koch2019a}.

A common assumption of mixed-dimensional models is that the radial scale of network tubes $R$
is much smaller than the dimension $\meas{\Omega}$ of the domain of interest $\Omega$. More specifically
for network systems, $R$ has to be much smaller than the average distance to the closest neighboring segment in the network.
While this precondition may be clearly satisfied in some cases (e.g. simple injection and extraction wells in large
distance to each other), it is less clear in others (blood capillaries with \SIrange{3}{8}{\micro\m}
vessel radius with average distances of \SI{50}{\micro\m}
in a microvascular network occupying about \SIrange{2}{3}{\percent} of the tissue volume).
Based on this main assumption, a couple of arguments are derived leading to the simplification of the model
equations. To allow for simple meshing procedure independent of the network domain,
the three-dimensional domain is extended to also cover the
volume occupied by the tubular network geometry~\citep{Laurino2019}.
This means the volume of the three-dimensional
domain is overestimated. Moreover, the tubes do not pose any resistance to flow since the physical
presence of the tubes is removed. Finally, since these mixed-dimension models
are usually derived for infinite cylindrical segments, some error is involved at
bifurcations by assuming finite cylindrical segments.
In this work, we want to investigate some of these assumptions numerically in more detail
than previously presented.

The mentioned mixed-dimensional models have been derived in the context of linear elliptic PDEs with line source terms.
Extensions to time-dependent problems and nonlinear problems have been analyzed to a much less extent.
Mixed-dimension methods are for example used to describe root water uptake from soil, in which case the
soil is described by the Richards equation---a strongly nonlinear PDE.
Although simple mixed-dimension methods have been used for more than two decades root water uptake simulations~\citep{Doussan1998},
the proper grid resolution required to accurately solve the model equations in dry soils
is rarely considered in the literature~\citep{Schroeder2009grid,Beudez2013}.
It is known that a coarse grid resolution in the soil domain may not accurately approximate local pressure gradients
and models for the root water uptake flux correction have been developed~\citep{Mai2019,Schroeder2008localsoil,Schroeder2009implmicro}.
To the best of our knowledge, the coupled root water uptake problem has not been rigorously analyzed mathematically.
The estimation of discretization errors and possible errors in the model reduction are yet to be better understood.
In this work we present a model technique to investigate such errors numerically.

In previous works on tissue perfusion models~\citep{dangelo2008}
and in more general mathematical works~\citep{koeppl2016},
it has been found that mixed-dimension embedded schemes
exhibit sub-optimal convergence rates, if the local discretization length $h$ is
larger than the radius $R$ of the embedded vessels.
While convergence rates alone are inconclusive about the error
at a given practical discretization length, the results in~\citep{koeppl2016,koeppl2018,Koch2019a}
indicate that in order to achieve sufficiently accurate numerical results,
the discretization length in the embedding bulk domain has to be chosen in the order of
the network tube radii or smaller. This is in stark contrast to typical grid resolutions
in root water uptake simulations, where soil cells
are routinely chosen an order of magnitude larger than the root radius~\citep{Schroeder2009grid,Leitner2014b}.
Several techniques to relax this
discretization length restriction have been discussed in the context of linear stationary
elliptic mixed-dimensional equations~\citep{Gjerde2018,Koch2019a,Koch2019bwell}.

In this work, we present a method where the tubular network is described by a network of line segments
with a given radius function as common in mixed-dimensional models. However, we explicitly resolve the
interface of the tubular network with the computational mesh describing the embedding bulk domain.
The new interface-resolving method developed subsequently can be considered
a reference method for the comparison of efficient mixed-dimension embedded schemes
based on implicit or reduced interface concepts.

For reference, we mention that in root-soil interaction simulations,
the root-soil interface has been explicitly resolved based on imaging data in a recent work by~\citet{Daly2018}.
However, only flow in the soil is simulated and the flow field is not coupled
to the flow field in the root xylem. Consequently, it is necessary to specify boundary conditions on the root-soil interface.
For the subsequently introduced method, the state of the root-soil interface is part of the solution.
Finally, we briefly introduced the new method in \citep{Schnepf2019benchmark}, where it is suggested
for the purpose of providing a reference solution in a benchmark study for root water uptake simulators.
In this work, we describe and analyse the method in more detail.

The new numerical method is derived in \cref{sec:projintro} (mathematical model)
and \cref{sec:disc} (discretization aspects) and
then applied in several numerical cases in \cref{sec:numeric}.
A grid convergence study in \cref{sec:onepbenchmark} shows
that the method is more accurate than other mixed-dimension methods for similar mesh sizes.
We compare the new method with previously published methods for examples for numerical test cases
of tissue perfusion and root water uptake in \cref{sec:tissue,sec:root}.

\section{Mixed-dimension method with resolved interface}
\label{sec:projintro}

\begin{figure}[htb!]
	\centering
	\includegraphics[width=0.6\textwidth]{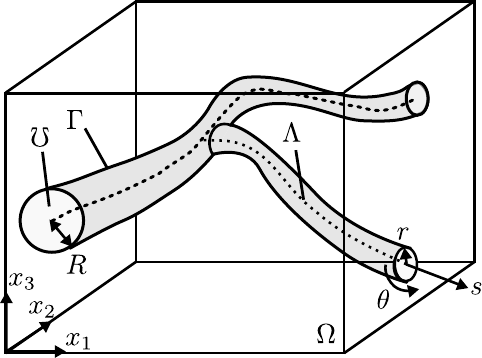}
	\caption{\textbf{Embedded tubular network system.} A tubular network structure
  with centerline skeleton $\Lambda$
	is embedded into the bulk domain $\Omega$. The surface of the tubes $\Gamma$ (dark grey) forms an internal boundary
  for $\Omega$, $\Gamma \subset \Omega \cap \mho$. Both domains are equipped
	with (local) coordinate systems. $R = R(s)$ denotes the equivalent local radius of the tube.}
	\label{fig:network}
\end{figure}

The tubular segments in network systems like plant roots or capillary blood vessels are usually
much smaller in radial extent than in axial extent, $R \ll L$. Often, it is therefore a good assumption
to neglect radial variations and work with cross-section averaged quantities and one-dimensional models
that describe the change of e.g. average pressure, temperature, concentration, etc. along the centerline axis~\citep{Koch2020PhDthesis}.
In this work we will assume that this one-dimensional description is sufficiently accurate and exploit this fact
by not resolving the network structure with a fully-resolved three-dimensional computational mesh which would
result in problems of intractable size. Moreover, we assume that any membrane separating the internal highly conducting space
of the tube (e.g. blood vessel lumen, root xylem) from the bulk domain can be described as a two-dimensional
sharp interface $\Gamma$.

In the following, we will exemplarily consider the case of root water
uptake.
For details on the mathematical modeling of root water uptake with three-dimensional root architectures,
we refer to the literature~\citep{Doussan1998,Javaux2008,Roose2008,Dunbabin2013,Koch2018a,Schnepf2019benchmark}.
Fluid flow in the root xylem---a structure that can be imagined as a bundle of tubes located in the center
of the root and transporting fluid in axial direction upwards toward the plant leaves---can be described by
\begin{equation}
\label{eq:xylem}
  - {\partial_s}\left( K_\text{ax} {\partial_s} p_{\soned} \right) = -q \quad \text{on} \quad \Lambda,
\end{equation}
with some boundary conditions on $\partial \Lambda$,
where $s$ denotes the local axial coordinate, $K_\text{ax}$ (in \si{\m\tothe{4}\per\pascal\per\s}) is
the axial root xylem conductivity, $p_{\soned}$ is the root xylem pressure (in \si{\pascal}), and $q$ (in \si{\square\m\per\s})
is a source term modeling fluid exchange with the embedding soil domain and depends on both $p_{\soned}$
and the soil pressure $p_{\sthreed}$ on the interface $\Gamma$.

Water flow in the soil is described by the Richards equation,
\begin{align}
\label{eq:richards}
	-\div \left(\frac{k_r(p_{\sthreed})}{\mu} K \grad p_{\sthreed}  + \rho\vec{g} \right)  &= 0 & \text{in} \quad \Omega, \\
\label{eq:coupling}
  -\left(\frac{k_r(p_{\sthreed})}{\mu} K \grad p_{\sthreed} + \rho\vec{g}\right) \cdot \vec{n}_\Gamma
  &= K_\text{r} \left[ p_{\sthreed}(\vec{x}_\Gamma) - p_{\soned}(\Pi\vec{x}_\Gamma) \right] & \text{on} \quad \Gamma,
\end{align}
with suitable boundary conditions prescribed on $\partial\Omega\setminus\Gamma$.
In \cref{eq:richards}, $\mu$ is the dynamic fluid viscosity (in \si{\pascal\s}),
$\rho$ is the fluid density (in \si{\kg\per\cubic\m}),
$k_r$ denotes the dimensionless relative permeability,
and $K$ the intrinsic permeability of the soil (in \si{\square\m}).
The relative permeability is a nonlinear function of $p_{\sthreed}$, e.g. modeled
by the well-known Van Genuchten-Mualem model~\citep{mualem1976,van1980closed}.
In \cref{eq:coupling}, $\vec{n}_\Gamma$
is an outward-pointing (with respect to $\Omega$) unit normal on $\Gamma$
and $K_r$ is the radial root conductivity (in \si{\m\per\pascal\per\s}).
Finally, $\Pi\bullet$ is a surjective projection operator that maps any point $\vec{x}_\Gamma$ on $\Gamma$
to a corresponding point $\hat{s}$ on $\Lambda$, given a parameterization of $\Lambda$ in terms of $\hat{s}$.

In order to obtain a mass conservative coupling scheme, we need to define the source term $q$ in \cref{eq:xylem}.
To this end, we denote with $\Lambda_\varsigma \subseteq \Lambda$ some compact subset of $\Lambda$ and with
$\Gamma_\varsigma = \lbrace \vec{x}_\Gamma \in \Gamma \mid \Pi \vec{x}_\Gamma \in \Lambda_\varsigma \rbrace$
the corresponding set of surface points on $\Gamma$. Then, the coupling condition is given by
\begin{equation}
\begin{split}
  \int_{\Lambda_\varsigma}\! q \,\text{d}s
    &= \int_{\Gamma_\varsigma}\! \left(\frac{k_r(p_{\sthreed})}{\mu} K \grad p_{\sthreed} + \rho\vec{g}\right) \cdot \vec{n}_\Gamma \,\text{d}\gamma \\
    &= \int_{\Gamma_\varsigma}\! -K_\text{r} \left[ p_{\sthreed}(\vec{x}_\Gamma) - p_{\soned}(\Pi\vec{x}_\Gamma) \right] \,\text{d}\gamma,
\end{split}
\end{equation}
given some suitable parameterization of $\Gamma_\varsigma$ in terms of $\gamma$.

\subsection{Practical geometry parameterization}
\label{sec:parametr}

\begin{figure}[!hbtp]
  \includegraphics[width=1.0\textwidth]{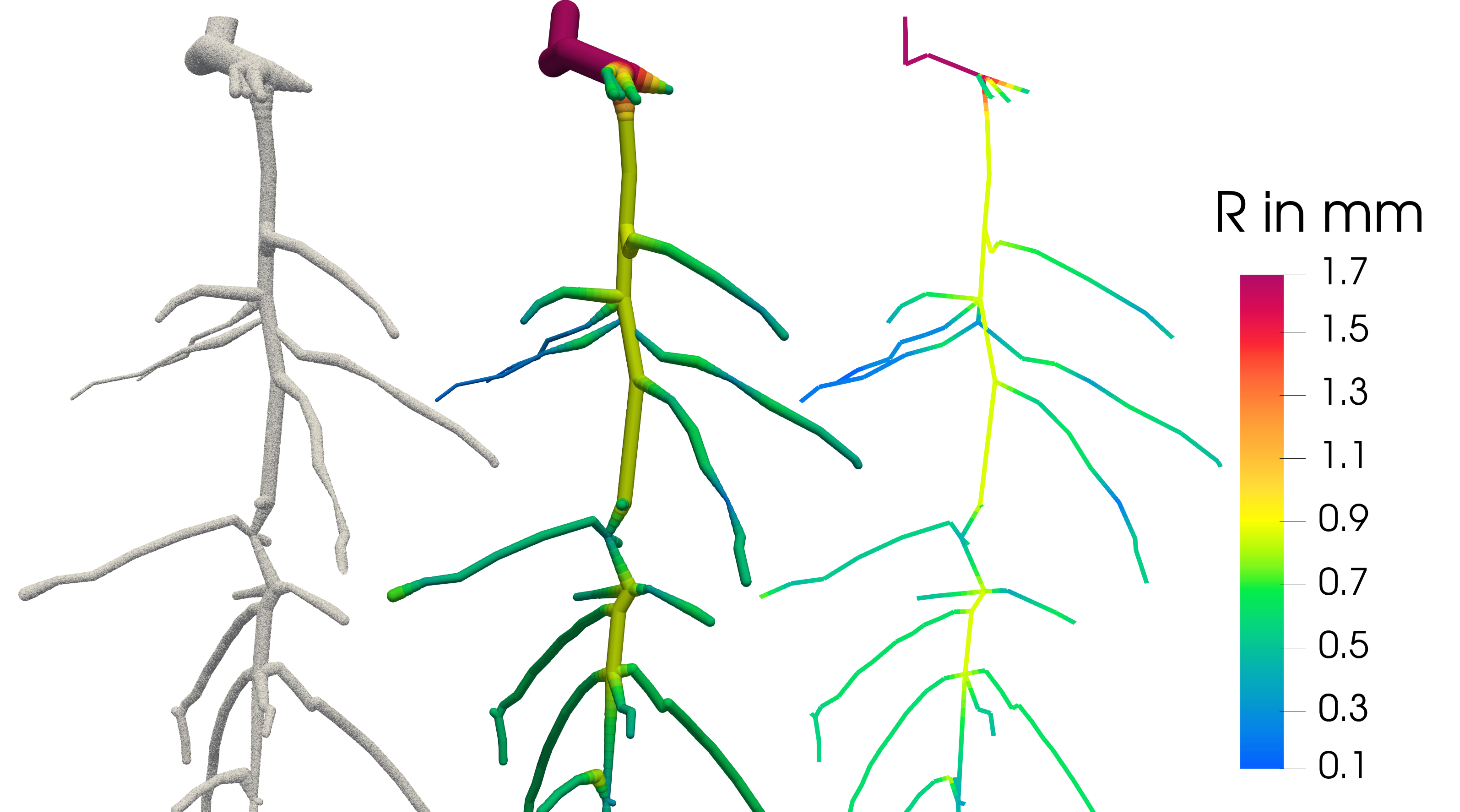}%
  \caption{\textbf{Three-dimensional representations of a segmented root architecture (lupine).} From right to left: graph representation with segment-wise continuous radius field,
  implicit geometrical representation as the sum of capsules, and discrete representation of root-soil interface
  as the surface facets of a tetrahedron mesh of the soil domain (generated with CGAL~\citep{CGAL2019}).}
  \label{fig:root_projection}
\end{figure}

The interface $\Gamma$ between network and bulk domain may often be given by some implicit description
in form of a continuous or discrete level set function, e.g. obtained from imaging data. From such
an implicit description it is possible to generate surface triangulations \citep{Boissonnat2005}, and extract
center-lines, for example based on the medial axis transformation~\citep{Lindquist1996,Antiga2008}.
Nevertheless, there is in general no unique choice for the mapping $\Pi$.

Since our method is targeted at the creation of verification tests for reduced methods
without explicit interface resolution, we simplify the geometrical description as follows.
The root network center-lines $\Lambda$ are approximated by $\Lambda_h$, a set of
linear root segment center-lines $\Lambda_i$ defined by two points $\vec{q}_i$, $\vec{p}_i$ and parametrized by
\begin{equation}
  \label{eq:line_param}
  \tilde{\vec{x}}(\tilde{s}_i) := \vec{p}_i + \tilde{s}_i \vec{m}_i, \quad \vec{m}_i=\vec{q}_i-\vec{p}_i, \quad \text{for}\quad \tilde{s}_i\in[0,1].
\end{equation}
Moreover, associated which each segment is a continuous radius function $R_i(\tilde{s}_i)$
which is often---but not necessarily---constant per segment but varies from segment to segment.
From this representation, we implicitly define a three-dimensional network representation
by the signed distance functions (SDFs),
\begin{equation}
\begin{split}
  \label{eq:implicitdomains}
  d_{\Lambda_h}(\vec{x}) &:= \min\limits_{\Lambda_i \in \Lambda_h} d_{\Lambda_i}(\vec{x}), \quad \vec{x} \in \mathbb{R}^3,\\
  d_{\Lambda_i}(\vec{x}) &:= \norm{\tilde{\vec{x}}(\mathbb{P}_i(\vec{x})) - \vec{x}} -R_i(\mathbb{P}_i(\vec{x})), \\
  & \text{where} \quad \hat{s}_i := \mathbb{P}_i(\vec{x}) = \max\left\lbrace 0, \min\left\lbrace\frac{(\vec{x} - \vec{p}_i)\cdot \vec{m}_i}{\lVert \vec{m}_i \rVert^2_2}, 1 \right\rbrace\right\rbrace.
\end{split}
\end{equation}
This parameterization allows for the convenient definition of the operator $\Pi$ to yield
the position $\hat{s}_i$ on the segment $\Lambda_i$ with minimal $d_{\Lambda_i}(\vec{x})$.
\begin{remark}
The SDFs $d_{\Lambda_i}$ describe capsules with radius function $R_i(\tilde{s}_i)$ around the line $\Lambda_i$
and the SDF $d_{\Lambda_h}$ describes the union of all such capsules.
Then every point $\vec{x}$ is inside the root, if $d_{\Lambda_h}(\vec{x}) < 0$, or in the soil or outside the domain, if $d_{\Lambda_h}(\vec{x}) > 0$.
Consequently, the root-soil surface is given by the zero level set $\Gamma = \lbrace \vec{x} \,\vert\, d_{\Lambda_h}(\vec{x}) = 0 \rbrace$.
\end{remark}

In this work, we use the meshing capabilities of the C++ geometry library CGAL~\citep{CGAL2019}
to generate computational grids for the bulk domain from such an implicit description.
These grid explicitly resolve the bulk-network interface.
An exemplary root network, the three-dimensional representation as a union of capsules, and a triangulated representation of $\Gamma$
is shown in~\cref{fig:root_projection}.

We note that the surface implied by the zero level set of \cref{eq:implicitdomains}
is only piecewise differentiable due to the possible discontinuity of $R$ between segments.
If necessary, the surface's smoothness can be improved by using a smooth minimum function for $d_{\Lambda_h}$, such as
\begin{equation}
\operatorname{smin}(a, b, k) := \min\{a, b\} - \frac{1}{6} h^3 k, \quad h = \frac{1}{k}\max\lbrace k - \vert a-b \vert, 0\rbrace,
\end{equation}
rendering the surface function twice differentiable ($C^2$)~\citep{smoothmin2013}.
The parameters $a$ and $b$ are signed distances
and $k > 0$ is a smoothing parameter (with units \si{\m})
which is to be chosen in the order of magnitude of the dimensions of the objects merged.
In the following, we do not perform such smoothing of the network surface,
and approximate the surface by the zero level set of the distance function~\cref{eq:implicitdomains}.

\subsection{Relation to implicit surface mixed-dimension methods}
\label{sec:implicit}
\begin{figure}[htb!]
  \includegraphics[width=1.0\textwidth]{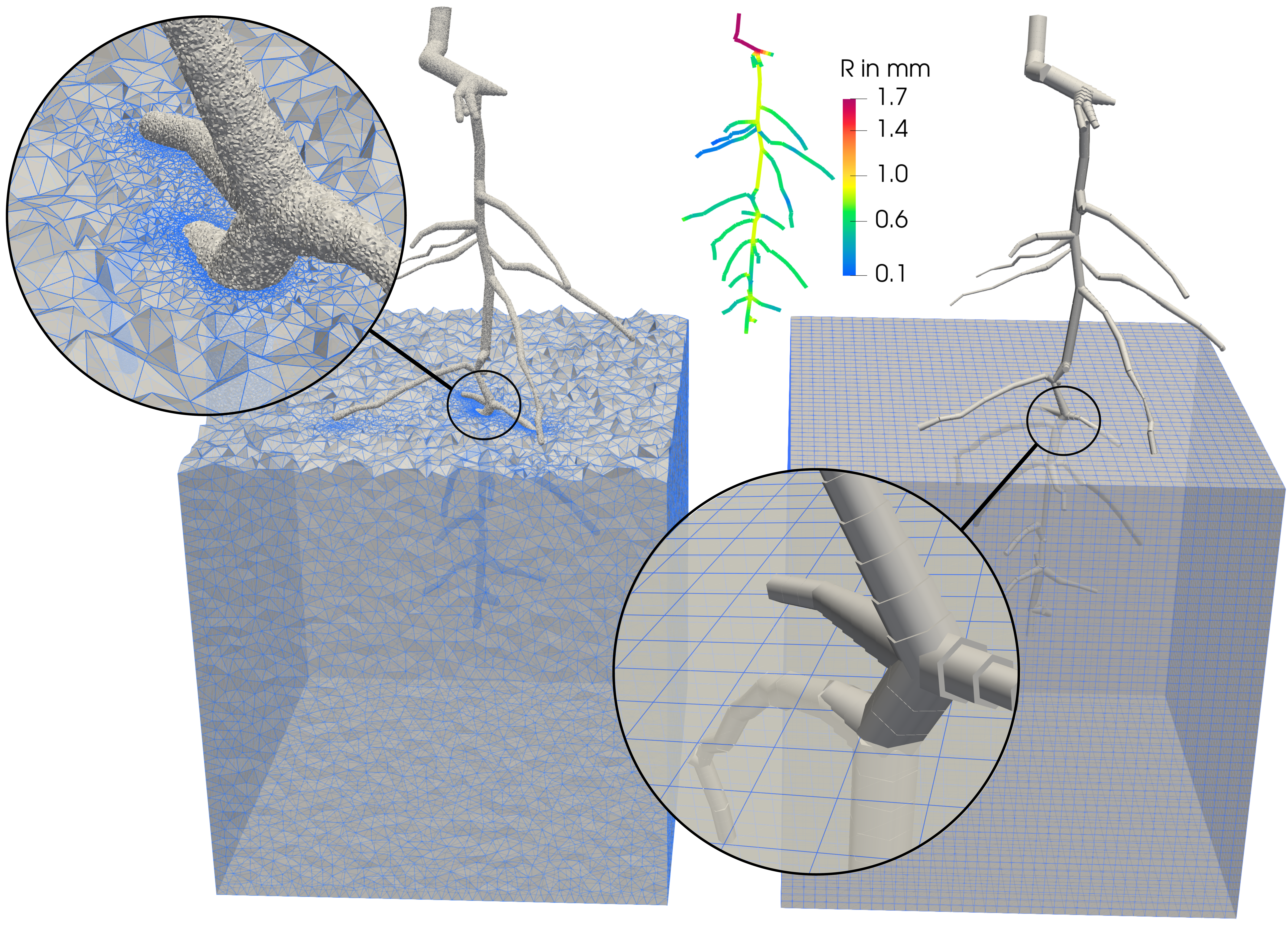}%
  \caption{\textbf{Comparison of two mixed-dimension discretization concepts.}
  A root system is embedded in a soil domain. Left, resolved-interface discretization. Coupling
  terms appear as boundary conditions for the 3D bulk domain. Right, implicit interface method with non-matching
  discretization. The bulk domain is extended to include the space occupied by the network (in its three-dimensional representation).
  Coupling terms for each segment appear as source terms restricted to a line~\citep{dangelo2008}, a local surface~\citep{koeppl2018}, or a local volume~\citep{Koch2019a},
  depending on the chosen method. This approach introduces additional model errors but implicit interface methods
  are expected to be computationally more efficient than the resolved-interface approach.}
  \label{fig:model_comparison}
\end{figure}

With the parameterization introduced in \cref{sec:parametr}, consider
a circular cross-section of radius $R_i$ of an infinite cylindrical tube.
Then, we can show that
\begin{align}
\label{eq:cylinder}
q(\tilde{s}_i) &= \int_0^{2\pi}\! -K_r \left( p_{\sthreed}(\vec{x}) - p_{\soned}(\Pi\vec{x}) \right) R_i\text{d}\theta \nonumber\\
        &= -2\pi R_i K_r \left( \frac{1}{2\pi}\int_0^{2\pi}\! p_{\sthreed}(\vec{x}) \,\text{d}\theta - \frac{1}{2\pi} \int_0^{2\pi}\! p_{\soned}(\Pi\vec{x}) \,\text{d}\theta \right) \nonumber\\
        &= -2\pi R_i K_r \left( \hat{p}^\bigcirc_{\sthreed} - p_{\soned}(\tilde{s}_i) \right),
\end{align}
where we used the fact that $p_{\soned}$ is independent of $\theta$,
and define the average pressure on the perimeter as
\begin{equation}
  \hat{p}^\bigcirc_{\sthreed} := \frac{1}{2\pi}\int_0^{2\pi}\! p_{\sthreed}(\vec{x}) \,\text{d}\theta.
\end{equation}
Hence, if the network is approximated by sufficiently long discrete cylinder segments
and exchange is assumed to only occur over the lateral surface of the cylinders, the source term can be formulated
solely in terms of quantities on the cross-sectional plane. Furthermore, assuming
that the network domain does not pose any resistance to flow in the bulk domain
and its volume is negligible, $\Omega$ is extended to include
the network domain, $\Omega^\text{ex} = \Omega \cup \mho$, and we arrive at a reduced model that can be written
in terms of a delta distribution restricting the bulk source term onto $\Gamma$, cf.~\citep{koeppl2018},
\begin{align}
	\label{eq:nonlineardiffusion-root}
	-\div \left(\frac{k_r(p_{\sthreed})}{\mu} K \grad p_{\sthreed} \right)  &= q\delta_\Gamma & \text{in} \quad \Omega^\text{ex}, \\
	\label{eq:1d-root}
	- {\partial_s}\left( K_\text{ax} {\partial_s} p_{\soned} \right) &= -q & \text{on} \quad \Lambda, \\
	\label{eq:source-root}
	q &= -2\pi R K_\text{r} (\hat{p}_{\sthreed}^\bigcirc - p_{\soned} ), \\
  \int_{\Omega^\text{ex}}\! q \delta_\Gamma \,\text{d}\vec{x} &= \int_{\Gamma}\! \frac{q}{2\pi R} \,\text{d}\gamma = \int_\Lambda\! q \,\text{d}s.
\end{align}
In the following, we refer to such formulations as \textit{implicit surface} methods as the exchange between bulk domain and network
is entirely formulated in terms of source terms in $\Omega^\text{ex}$ instead of boundary conditions
on $\Gamma$ and an explicit resolution of the bulk-network
interface by the computational mesh of the bulk domain is not necessary anymore.
However, we note that the interface $\Gamma$ still appears implicitly
in form of the delta distribution. \Cref{fig:model_comparison} compares the meshes
used for resolved-interface descriptions and implicit interface description in non-matching discretization schemes.

In \citep{d2007multiscale}, the author suggests
to use $\delta_\Lambda$ instead of $\delta_\Gamma$ in \cref{eq:nonlineardiffusion-root}, i.e. the source
term is restricted to a line in $\Omega^\text{ex}$. This formulation results in pressure solutions with feature
singularities on $\Lambda$ and are difficult to approximate by numerical schemes. In \citep{Koch2019a}, the authors suggest to replace the surface source
term by a volume source term using a distribution kernel in combination with a local reconstruction scheme of the interface
pressure $\hat{p}_{\sthreed}^\bigcirc$. This technique allows to decouple the discretization length from the tube radius,
but the local reconstruction scheme has only been investigated for linear problems so far (corresponding to a constant relative permeability $k_r$ in \cref{eq:nonlineardiffusion-root}).
In \cref{sec:numeric}, we compare the new resolved-interface method with implicit interface methods in numerical experiments.
To this end, we follow the terminology of \citep{Koch2019a} and refer with \textsc{css} (\textit{cylinder surface source}) to the method due to \citep{koeppl2018} using
formulation \cref{eq:nonlineardiffusion-root,eq:1d-root,eq:source-root}. We refer with \textsc{ls} (\textit{line source}) to the method due to \citep{d2007multiscale}
where $\delta_\Gamma$ is replaced by $\delta_\Lambda$, and with \textsc{ds} (\textit{distributed source}) to the method due to \citep{Koch2019a},
where $\delta_\Gamma$ is replaced by a volumetric distribution kernel and the source term $q$ is computed based on a local
reconstruction scheme.

\subsection{Integration of the coupling term}
\label{sec:disc}

In the discrete setting, the root and the soil domain, $\Lambda$ and $\Omega$
are partitioned into a finite number of grid cells such that
$\Lambda_h = \bigcup K_\Lambda$ and $\Omega_h = \bigcup K_\Omega$ are discrete mesh representations
of $\Lambda$ and $\Omega$ with the cells $K_\Lambda$ and $K_\Omega$.
The computational grids can be chosen independently.
A part of the boundary of $\Omega$ explicitly resolves the root-soil interface $\Gamma$ and
$\Gamma_h$ denotes the set of cell facets on the interface.
Since the interface is explicitly described by $\Gamma_h$, the
the coupling conditions, \cref{eq:coupling}, can be directly evaluated by
numerically approximating the surface integrals. However, in the discrete setting,
the approximation of $p_{\sthreed}$ is typically only piecewise differentiable.
In this work, we consider piecewise linear functions. The
source term $q$ needs to be integrated over a control volume $K_\Lambda$ which may involve integration
over several interface facets. For this purpose, we suggest an algorithm based on virtual local refinement
of the interface facets to accurately capture the surface integration area element
associated with the integration over $K_\Lambda$. The algorithm is given as pseudo-code in \cref{algo}.
In brief, we map the corners of a surface triangle with $\Pi$ and evaluate if the mapped points
are contained in different network control volumes. If so, the triangle is virtually refined and
the procedure is repeated recursively until all corners map to the same control volumes, or some
maximum refinement level is reached. We add only one integration point per surface
triangle and coupled network control volume at the centroid of the union of coupled sub-triangles.
For $p_{\sthreed}$ and $p_{\soned}$ being piecewise linear functions,
integrating with the mid-point rule is exact.

\section{Numerical results and discussion}
\label{sec:numeric}

In this section, we compare the introduced explicit interface method
with previously suggested implicit interface methods in three cases.
We use the abbreviations \textsc{ls}, \textsc{css}, \textsc{ds} introduced
in \cref{sec:implicit} for the implicit interface methods and abbreviate with
\textsc{ps} (\textit{projection source}) the explicit interface method.
In the first case, \cref{sec:onepbenchmark}, we show for a simple rotation-symmetric setup with a single
tubular inclusion that \textsc{ps} accurately approximates a given
analytical solution and verify that the surface integration scheme
proposed in \cref{sec:disc} is sufficiently accurate.
In the second case, \cref{sec:tissue}, we investigate errors introduced by implicit interface
methods at the example of tissue perfusion described by a linear elliptic
mixed-dimensional model. In the third case, \cref{sec:root},
we investigate differences between \textsc{ps} and \textsc{css} in
a root water uptake example described by a nonlinear elliptic mixed-dimension model
based on the Richards equations.

The three-dimensional bulk domains $\Omega$, $\Omega^\text{ex}$ and the network domain $\Lambda$
are spatially decomposed into the meshes $\Omega_h$, $\Omega_h^\text{ex}$ and $\Lambda_h$ consisting
of cells $K_\Omega \in \Omega_h$ and $K_\Lambda \in \Lambda_h$, respectively.
The discretization length computed as the maximal cell diameter is denoted by $h$.
We discretize the continuous equation in space using finite volume methods.
For structured Cartesian grids as well as for the network equations, we
use a cell-centered finite volume method (\textsc{fvm}) with a two-point flux approximation (\textsc{tpfa}), cf.~\cite{Koch2019a}.
When using (unstructured) tetrahedral meshes for the \textsc{ps} method,
or in the case of locally refined meshes for the \textsc{css} method in \cref{sec:root},
we use vertex-centered
finite volumes with linear basis functions
(also referred to as \textsc{box} method)~\citep{Hackbusch1989,HuberHelmig1999,Koch2019cdumux}.
This is because cell-centered \textsc{tpfa}-\textsc{fvm} are generally not consistent
on such meshes~\citep{Schneider2018}.
The resulting discrete system of equations is solved with Newton's method. (In case of a linear
model Newton's method converges in one step.)
The linearized system of equations within each Newton iteration,
is solved with a stabilized bi-conjugate gradient method using a block-diagonal preconditioner
based on incomplete LU-factorization, cf.~\citep{Koch2019a}.
All presented methods and simulations are implemented using the open-source software framework \dumux~\cite{Koch2019cdumux}
with the network grid implementation \texttt{dune-foamgrid}~\cite{foamgrid} for representing the embedded network domain.

\subsection{Mixed-dimension single phase flow}
\label{sec:onepbenchmark}

\begin{figure}[htb!]
  \includegraphics[width=1.0\textwidth]{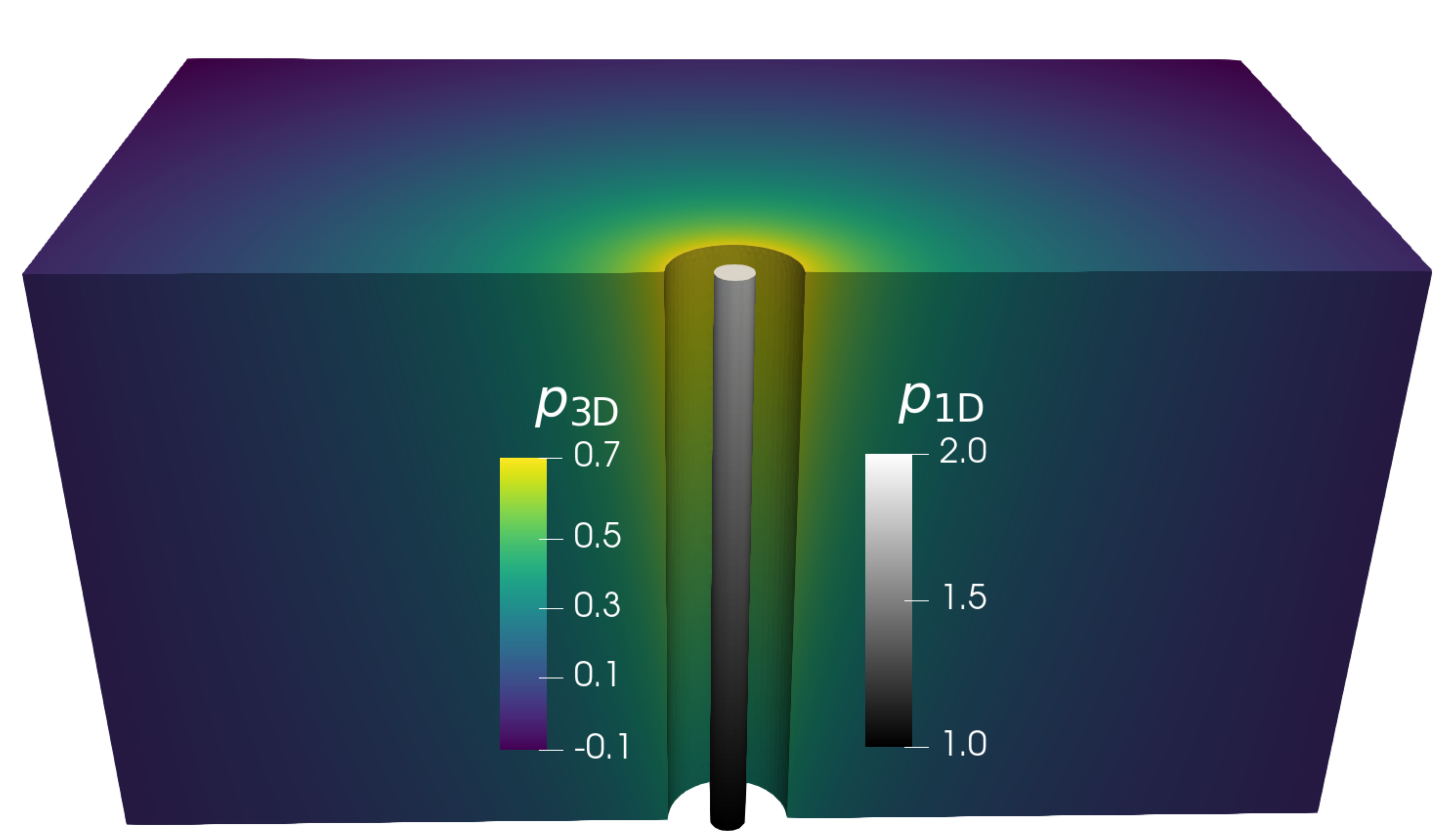}%
  \caption{\textbf{Reference solution for an embedded cylinder.}
  Cylindrical tube embedded in a box-shaped domain with dimensions $[-1,1]\times[-1,1]\times[-1,1]$.
  The domain is cut in half revealing the tube of radius $R = 0.03$ visualized with a reduced radius
  to make the interface visible. The numerical solutions $p_{\sthreed}^\textsc{ps}$,
  $p_{\soned}^\textsc{ps}$ computed on a fine grid with $h = 0.008$ are shown exemplarily
  but they are visually identical to a plot of the analytical solution.}
  \label{fig:cylinder_bench_plot}
\end{figure}
\begin{figure}[hbtp]
  \includegraphics[width=1.0\textwidth]{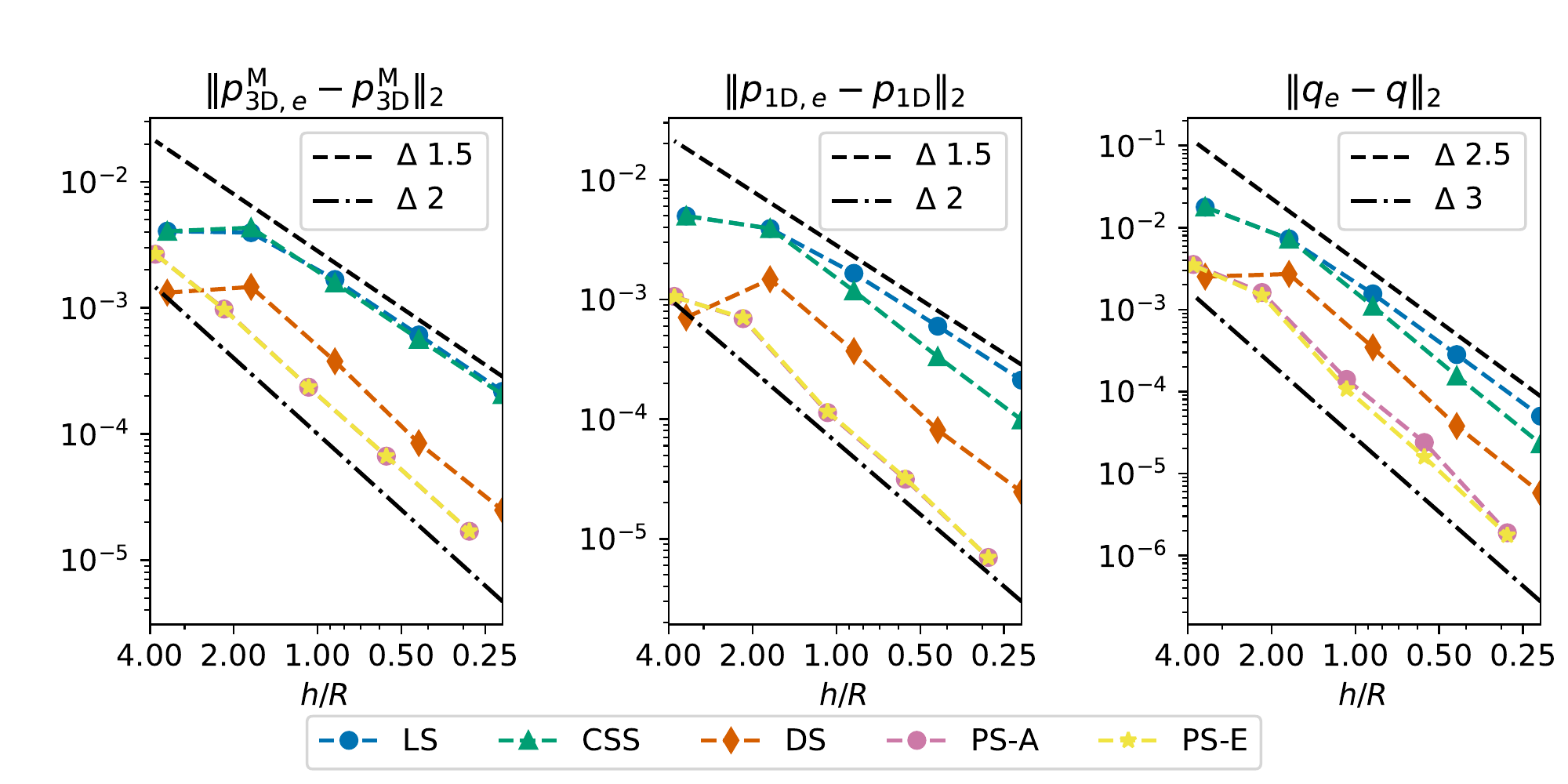}%
  \caption{Convergence rates over discretization length.}
  \label{fig:cylinder_bench}
\end{figure}

Let us consider a slightly simplified problem, adapted from \citet{d2007multiscale,Koch2019a},
(for simplicity we use the same symbols as previously introduced but
all unknowns and parameters are to be interpreted as dimensionless quantities,)
\begin{subequations}
\label{eq:kernel:problembeta}
\begin{align}
  -\partial_s \left(K_\text{ax} \partial_s p_\soned \right) &= -q \quad&& \text{in } \Lambda,\\
  -\div(\grad p_\sthreed) &= q\delta_\Lambda \quad&& \text{in } \Omega^\text{ex}, \\
  q &= - 2\pi R K_r (\hat{p}_{\sthreed}^\bigcirc - p_\soned),
\end{align}
\end{subequations}
with the domains $\Omega^\text{ex} = [-1,1]\times[-1,1]\times[-1,1]$ and $\Lambda = \{0\}\times\{0\}\times[0,1]$,
i.e the vessel center-line coincides with the $\xThree$-axis. The tube has radius $R$ and $\mho$
is given by the cylinder with center-line $\Lambda$, radius $R$ and unit length.
Recall that for this straight cylindrical tube case, due to the observation in \cref{eq:cylinder}, problem formulation \cref{eq:kernel:problembeta}
is equivalent to the formulation with boundary conditions on $\Gamma$, cf.~\cref{eq:xylem,eq:richards,eq:coupling} and
\begin{equation}
  \label{eq:cyl_benchmark_coupling}
  \int_\Lambda \! q \,\text{d}s = \int_{\Omega^\text{ex}} \! q\delta_\Lambda \,\text{d}x = \int_{\Gamma}\! -K_\text{r} \left[ p_{\sthreed}(\vec{x}_\Gamma) - p_{\soned}(\Pi\vec{x}_\Gamma) \right] \,\text{d}\gamma.
\end{equation}
Choosing the conductivities as
\begin{equation*}
K_\text{ax} = 1 + \xThree + \frac{1}{2}\xThree^2, \quad K_r = \left(2\pi R + R\ln {R}\right)^{-1},
\end{equation*}
the pressure solutions,
\begin{equation}
\label{eq:analytic_benchmark}
p_{\soned,\up{e}} = 1 + \xThree, \quad p_{\sthreed,\up{e}} = - \frac{1 + \xThree}{2\pi} \ln r, \quad r \geq R
\end{equation}
with $r = \sqrt{\xOne^2 + \xTwo^2}$, solve \cref{eq:kernel:problembeta} with matching boundary conditions.
From the analytical pressure solutions follows that $q_\up{e} = 1+\xThree$ is the analytical source term.

In this setting, we can directly compare the explicit interface method with
implicit interface methods.
The analytical solutions $p_{\soned,\up{e}}$ and $q_\up{e}$ are identical for all mentioned methods, cf.~\citep{Koch2019a}.
The exact pressure in the bulk domain, $p_{\sthreed,\up{e}}$, can be extended to all of $\Omega^\text{ex}$ and differs between the different methods
for $r < R$ (and $r < \varrho$ for the distributed source method (\textsc{ds}) of \citep{Koch2019a} where $\varrho$ is the radius of the distribution kernel),
but is identical for $r \geq R$ ($r \geq \varrho$ for \textsc{ds}).
The analytical solutions $p_{\sthreed,\up{e}}^{\textsc{m}} \in \Omega^\text{ex}$ for the methods $M \in \lbrace \textsc{ls}, \textsc{css}, \textsc{ds} \rbrace$ are given in \cref{app:benchmark}.

For the resolved interface method \textsc{ps}, in the case of straight cylindrical vessels,
integration of the source term $q_{K_\Lambda}$ over the discrete interface $\Gamma_h$ can be performed exactly.
To this end, we compute for every segment $K_\Lambda$ the intersections
of the space between the two planes implied the segment $K_\Lambda$ (the two planes through the end points
and normal to the segment) and all surface triangles $T \in \Gamma_h$. Effectively,
every $T$ is sub-triangulated such that every sub-triangle couples with exactly one $K_\Lambda$.
The boundary integral can be computed by the mid-point rule which is exact since the integrand is a linear function.
Intersections cannot be so easily computed in networks where the interface is only given as an
implicit function. Therefore, we suggested an approximate integration algorithm based on virtual refinement
in \cref{sec:disc}. For this particular test case, we implemented both approaches to verify
the accuracy of the latter. We denote the exact approach with \textsc{ps-e}
and the approximate approach with \textsc{ps-a}.

We solve \cref{eq:kernel:problembeta} with the methods
\textsc{ls}, \textsc{css}, \textsc{ds}, \textsc{ps-e} and \textsc{ps-a} and by prescribing
the analytical solutions as Dirichlet boundary conditions, except for the top and bottom sections of $\Omega$ or $\Omega^{ex}$ ($x_3 = 0$, $x_3 = 1$)
where we prescribe the normal derivative of the analytical solution as Neumann boundary condition.
Obviously, for the methods \textsc{ps-e} and \textsc{ps-a}, $\Gamma$ does not require boundary conditions
and fluxes over $\Gamma$ are computed by the coupling conditions, \cref{eq:cyl_benchmark_coupling}.
The numerical solutions $p^{\textsc{ps}}_\sthreed$ and $p_\soned$ for ${R} = 0.03$
are shown in~\cref{fig:cylinder_bench_plot}.

We compute pressure discretization errors in the normalized discrete norm
\begin{equation}
 \lVert p_\sthreed - p^{\textsc{m}}_{\sthreed,\up{e}} \rVert_2 := \frac{\left[ \sum_{\Omega_h} \vert K_\Omega \vert (p^{\textsc{m}}_{K_\Omega,\up{e}} - p_{K_\Omega})^2 \right]^{1/2}}{\sum_{\Omega_h} \vert K_\Omega \vert},
\end{equation}
where $p_{K_\Omega}$, $p_{K_\Omega,\up{e}}$ denote numerical and exact pressure evaluated at the center of a control volume
$K_\Omega$ and $\vert K_\Omega \vert$ its volume. The error for $p_\soned$ in $\Lambda_h$ is computed analogously.
The error in the source term $q$ is computed as
\begin{equation}
 \lVert q - q_{e} \rVert_2 = \frac{\left[\sum_{\Lambda_h} \vert K_\Lambda \vert (q_{K_\Lambda,\up{e}} - q_{K_\Lambda})^2\right]^{1/2}}{\sum_{\Lambda_h} \vert K_\Lambda \vert},
\end{equation}
where
\begin{equation}
  \label{eq:source_discrete}
 q_{K_\Lambda,\up{e}} = \int_{K_\Lambda} \!q_\up{e}\,\text{d}s \quad \text{and} \quad q_{K_\Lambda} = \int_{K_\Lambda}\! q \,\text{d}s.
\end{equation}
The maximum control volume size, $h$, is given by the maximum cell diameter in both domains.
We choose $h = h_\Omega$ such that $h_\Omega \approx h_\Lambda$. Both domains are uniformly refined.
The mesh for \textsc{ps} is remeshed so that the discrete interface $\Gamma_h$ approaches the real interface
$\Gamma$ with grid refinement.
Pressure and source error norms with grid refinement are shown in~\cref{fig:cylinder_bench}.

For sufficiently smooth solutions, the employed finite volume schemes
are expected to show a quadratic error decay of both pressures with grid refinement in the specified discrete norms.
However, $p_{\sthreed,\up{e}}^\textsc{ls}$ exhibits a singularity for all $\vec{x} \in \Lambda$
and $p_{\sthreed,\up{e}}^\textsc{css}$ has a kink on $\Gamma$. Therefore, the
convergence rate is reduced for these methods. Unfortunately, for the $\textsc{ls}$ method
the convergence order of $p_{\soned,\up{e}}^\textsc{ls}$ is affected by the reduced
convergence order of $p_{\sthreed,\up{e}}^\textsc{ls}$.

It is evident from the convergence results that for a given grid resolution the
\textsc{ps} method shows the smallest error of all presented methods. Furthermore,
the integration scheme suggested in \cref{sec:disc} (\textsc{ps-a}) is accurate enough and matches
the results with the exact integration formula (\textsc{ps-e}) well.
This motivates the conclusion that the newly introduced explicit interface method may serve as a reference for
implicit interface methods.

\subsection{Fluid perfusion of vascularized tissue}
\label{sec:tissue}

Fluid flow in the capillary blood vessels,
in the (fluid-filled) extra-vascular extra-cellular space (interstitium),
and fluid exchange between these compartments
can be described by linear mixed-dimensional
PDE systems~\citep{dangelo2008,cattaneo2014computational,Koch2019a}.
In this section, we consider the following model:
\begin{subequations}
  \label{eq:tissue-perfusion-problem}
\begin{align}
  \label{eq:blood}
    - \partial_s\left( K_\text{ax} \partial_s p_{\soned} \right) &= -q  &\text{on} \quad \Lambda,\\
  \label{eq:tissue}
    -\div \left(\frac{K}{\mu_I} \grad p_{\sthreed} \right)  &= 0 & \text{in} \quad \Omega, \\
  \label{eq:tissue_coupling}
    -\left(\frac{K}{\mu_I} \grad p_{\sthreed} \right)\cdot \vec{n}_\Gamma
    &= K_\text{r} \left[ p_{\sthreed}(\vec{x}_\Gamma) - p_{\soned}(\Pi\vec{x}_\Gamma) + \Delta \pi \right] & \text{on} \quad \Gamma, \\
  \label{eq:tisue_coupling_coupling}
    \int_{\Lambda_\varsigma}\! q \,\text{d}s
    &= \int_{\Gamma_\varsigma}\! -K_\text{r} \left[ p_{\sthreed}(\vec{x}_\Gamma) - p_{\soned}(\Pi\vec{x}_\Gamma) + \Delta \pi \right] \,\text{d}\gamma, &
\end{align}
\end{subequations}
where here $p_{\soned}$ denotes the blood pressure, $K_\text{ax} = \frac{\pi R^4}{8 \mu_B}$
is the axial conductivity with the apparent blood viscosity $\mu_B$, here taken as a constant;
$p_{\sthreed}$, $\mu_I$ denote the interstitial fluid pressure and viscosity,
$K$ is the intrinsic permeability of the interstitium, and $\Delta \pi$ is the
colloid osmotic pressure difference between both compartments, often assumed constant~\citep{levick1991}.
The corresponding implicit interface model
can be derived analogously to \cref{eq:nonlineardiffusion-root,eq:1d-root,eq:source-root}
and is discussed for various implicit interface methods in more detail in \citep{Koch2019a}.

We consider two scenarios. First, the fluid flow on a cross-sectional cut plane
through several parallel infinitely long vessels with different but constant pressures.
In this case we investigate the error involved in neglecting the vessel volume
and resistance in the bulk domain by extending $\Omega$ to $\Omega^\text{ex}$.
Second, we consider coupled fluid flow in and around
a small three-dimensional vessel network extracted from the rat brain.
In this case we investigate the error involved in approximating vessel
bifurcations by possibly overlapping cylinder segments as frequently done
in implicit interface methods.

\subsubsection{Effect of neglecting vessel resistance to bulk flow}
\label{sec:2d-multiple}

In this section, we show with a numerical example comparing the explicit interface
\textsc{ps} method with implicit interface methods that neglecting the resistance of the
vessel to bulk flow introduces some error in the bulk pressure field and the computed
exchange source term.
However, this error is likely small and may be neglected in practical simulations.

Consider a scenario with several parallel vessels of different radius
and constant but different vessel pressures. For this particular case,
the system \cref{eq:tissue-perfusion-problem} can be reduced
to two dimensions, since all cross-sectional planes have identical solutions.
However, for code verification purposes such a scenario can still
be simulated as a three-dimensional problem. To this end, we restrict the meshes for
bulk and vessel domain to a single cell in the axial direction. The vessel pressure is fixed
(Dirichlet boundary conditions) and the top and bottom plane (axial cross-sectional plane)
are assigned no-flow boundary conditions (homogeneous Neumann boundary conditions).

\begin{remark}
It is known that for the particular case of parallel vessels and constant vessel pressures \citep{koeppl2018,Koch2019a},
the solution obtained with one of the implicit interface methods (\textsc{ls}, \textsc{css} or \textsc{ds})
converges to a solution $p_{\sthreed,e} \in \mathbb{R}^2$ on each cross-sectional plane
that can be written as the superposition of fundamental solutions
and a harmonic function $H$ chosen to satisfy given boundary conditions on $\partial\Omega \setminus \Gamma$,
\begin{equation}
\label{eq:superpos}
  p_{\sthreed,e} = H + \sum_i^N \frac{q_i}{2\pi}ln{\frac{\lVert \vec{x}_i - \vec{x}\rVert_2}{R_i}}, \quad q_i = -2\pi R K_\text{r} (p_{\sthreed,e,i}^{\bigcirc} - p_{\soned,e,i} ),
\end{equation}
where $N$ is the number of vessels, $\vec{x}_i$ is the centerline position and $R_i$ the radius of vessel $i$,
$p_{\soned,e,i}$ denotes the given vessel pressure of vessel $i$ and $p_{\sthreed,e,i}^\bigcirc$ the average
bulk pressure on the perimeter of vessel $i$. Taking the average of \cref{eq:superpos} over every vessel perimeter
results in a system of $N$ equations with $N$ unknown $p_{\sthreed,e,i}^{\bigcirc}$. The system can be solved numerically
to obtain a simple expression for $p_{\sthreed,e}$ in terms of known $q_i$, cf.~\citep{Koch2019a}.
For continuations of the function to $\Omega^\text{ex}$ in consistency with the respective method,
see \citep{Koch2019a}.
\end{remark}

On the other hand, the \textsc{ps} method converges
to a different (but physically more sensible) solution since the vessel volume
is actually excluded from the domain and the vessels therefore act as virtually impermeable
(due to the low permeability of the vessel wall) obstacles to flow in the bulk domain.
This vessel resistance is neglected in the derivation of implicit interface methods
when the extra-vascular domain is extended to $\Omega^\text{ex}$ neglecting the vessel volume.

\begin{figure}[htb!]
  \includegraphics[width=1.0\textwidth]{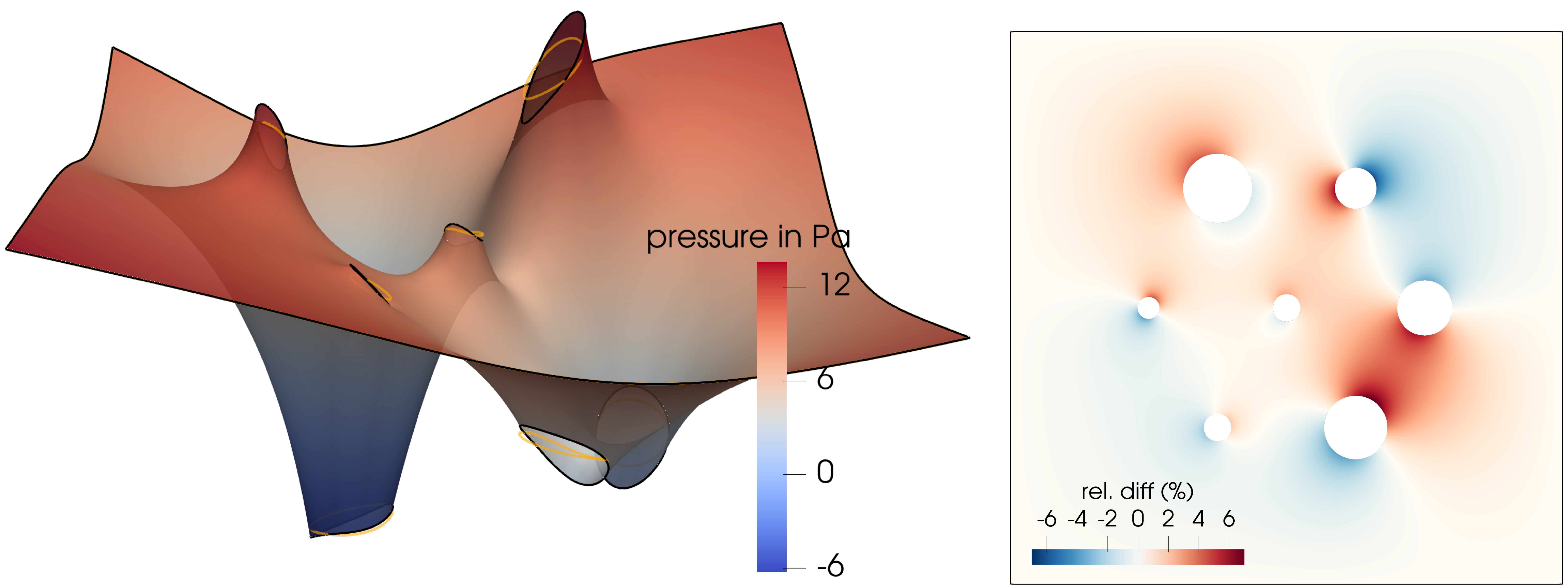}%
  \caption{\textbf{Effect of neglecting vessel resistance to bulk flow.}
  Pressure distribution in the interstitial space on a cross-sectional plane
  around seven parallel circular vessels with different but constant pressures. Fluid
  flow is driven by the pressure difference between vessels (holes in the visualization)
  and interstitium. Some vessels
  produce fluid (arterial end) and some absorb fluid (venous end) from the interstitium.
  Left, surface plot of the pressure distribution resulting from the resolved interface method.
  Orange circles show the pressure distribution on the vessel surface when neglecting
  the vessel volumes (and the associated resistance to bulk flow) in implicit interface methods.
  Right, pressure difference in the interstitium between the numerical solution of the resolved interface method
  and the analytical solution corresponding to implicit interface methods. The difference is relative to
  the maximum absolute bulk pressure.}
  \label{fig:multiple2d}
\end{figure}

We consider a scenario with $7$ parallel vessels. The case is chosen
such that the distances between vessels are unusually small and pressure differences
between neighboring vessel are large. For this setup, pressure gradients in the
bulk domain are strongly influenced by neighboring vessels. Therefore, possible
differences between implicit and explicit interface schemes are expected to be particularly large.
For simplicity, we here choose $H = 0$. The other parameters and the computed
$q_i$ are given in \cref{tab:tissue,tab:2d_multiple}.
\Cref{fig:multiple2d} shows a comparison of
the numerical pressure solution for the \textsc{ps} method in comparison with
the analytical solution \cref{eq:superpos} for implicit interface methods.
Dirichlet boundary conditions on
the outer boundary fix the solution to \cref{eq:superpos}.

As evident in \cref{fig:multiple2d} the local bulk pressure differs significantly
close to the vessel surface (up to \SI{7}{\percent}). However, the difference
diminishes rapidly in some distance to the vessel. Moreover, we show the bulk
pressure distribution on the vessel interface for both cases in \cref{fig:multiple2d}(left).
For the \textsc{ps} method the pressure varies significantly. With respect to the bulk
flow direction the interface pressure is higher upstream and lower downstream due the
resistance posed by the vessel. This variance is considerably reduced in the
implicit interface case where this resistance is neglected.

We recall that in the given scenario, the source terms $q_i$
depend on the average interface pressure for both methods. Remarkably,
the differences in $p_{\sthreed,i}^{\bigcirc}$ are much lower than point-wise differences.
The largest difference (relative to the maximum bulk pressure) is found for vessel $1$
with \SI{1.2}{\percent} and the smallest for vessel $3$ with \SI{0.1}{\percent}.
Therefore, although bulk pressure may differ significantly at the interface, the source
term are estimated relatively accurate. Another important aspect leading to even smaller
differences in the source term is the fact that the bulk-vessel pressure drop is
usually dominated by the pressure drop over the vessel wall membrane. Therefore, possible
errors in $p_{\sthreed,i}^{\bigcirc}$ are not categorically visible in $q_i$.
Interestingly, the largest difference in $q_i$ (relative to the maximum absolute source term)
is found to be \SI{0.02}{\percent} for vessel $1$, while the smallest difference
is \SI{0.0001}{\percent} for vessel $3$.

We conclude that such differences are negligible in the vast majority
of applications where usually exchange fluxes and conditions in some distance to the vessel
(e.g. oxygen concentration in a diffusion problem)
are of particular interest.

As a final remark, we want to mention that the resistance of the embedded network
to flow in the bulk can be incorporated in implicit interface methods by assigning a
low permeability to cells which are fully contained in $\mho = \Omega^\text{ex} \setminus \Omega$.
However, this may lead to ill-conditioned systems if this permeability value is chosen too low.
Such cells (and associated degrees of freedom) can also be entirely removed from the mesh.
However then, the efficiency of structured Cartesian grids might not be fully exploitable.
In both cases, the resolution of the 3D mesh needs to be fine enough to actually
resolve the vessel geometry.

In our experience, local parameter adjustment or cell removal
is not necessary to obtain sufficiently accurate results
with implicit interface methods. As suggested by the scenario in this section,
the introduced error by neglecting vessel resistance to bulk flow is small.
(This also explains why the \textsc{ds} method~\citep{Koch2019a} is able to
produce accurate results despite coarse grid resolution which are achieved by an interface pressure reconstruction
technique that necessitates the negligence of vessel resistance to bulk flow.)
Furthermore, as we will demonstrate in the subsequent sections,
other types of model and discretization errors usually dominate.

\subsubsection{Effect of bifurcation geometry approximations}
\label{sec:bifu}

\begin{figure}[htb!]
  \includegraphics[width=1.0\textwidth]{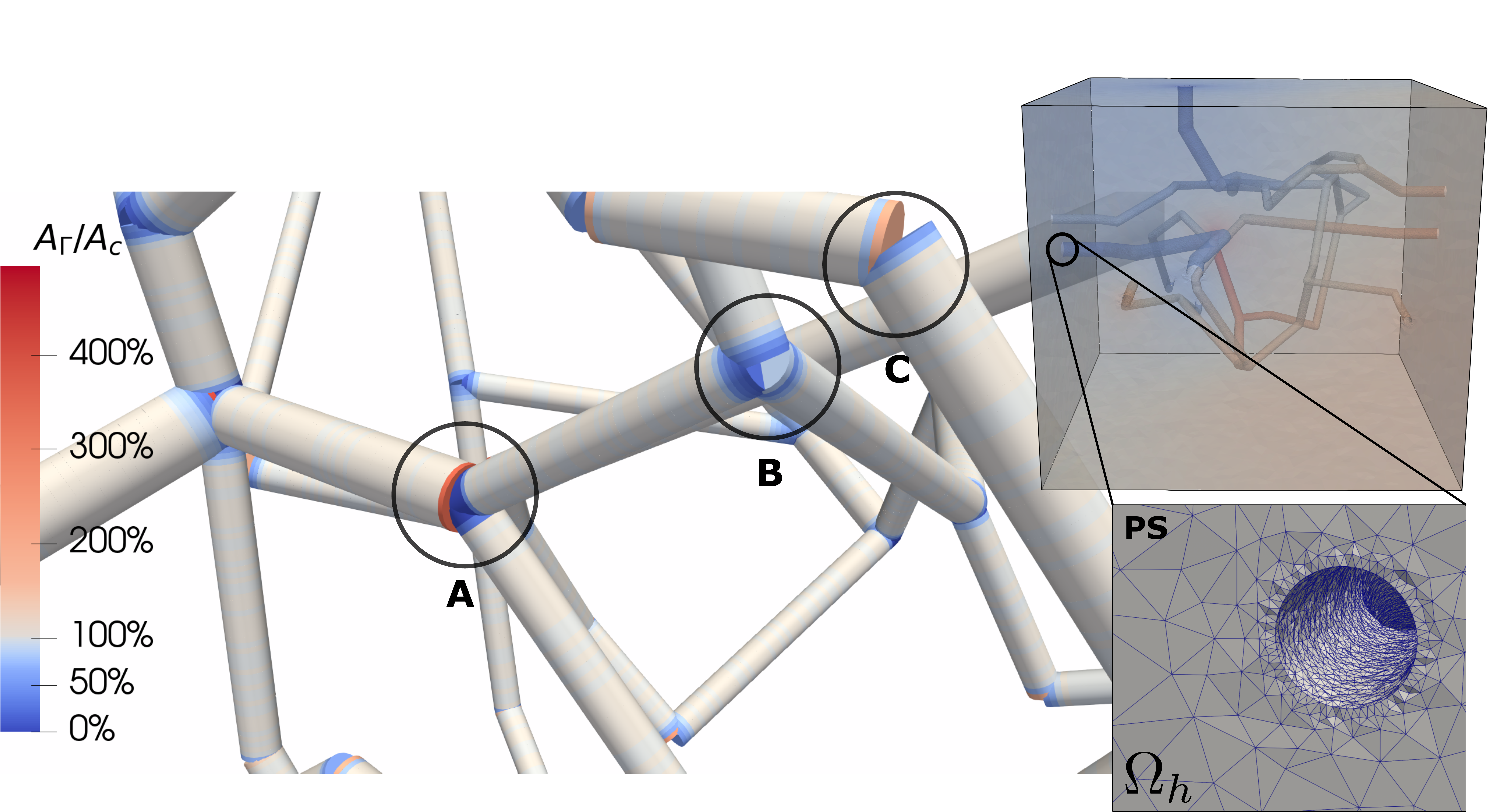}%
  \caption{\textbf{Interfacial area approximation in by discrete meshes.} Left, ratio of $A_\Gamma$,
  the surface area approximated by the interface-resolving mesh used for the \textsc{ps} method,
  and the surface area of a cylinder $A_c = 2\pi R L$, where $R$ and $L$ are radius and length
  of a network cell $K_\Lambda$. White color represents equal areas ($A_\Gamma/A_c = \SI{100}{\percent}$).
  $A_c$ is commonly used for implicit surface methods.
  At bifurcations (A, B) and kinks (C) cylinders of neighboring vessel may overlap leading
  to large local differences in the surface area with respect to an explicit surface representation.
  Right, boundary and interface faces of the mesh $\Omega_h$. Zoom-in shows the locally
  refined mesh around an exemplary vessel.}
  \label{fig:bifurcations}
\end{figure}

In this section, we solve a fluid perfusion problem in a tissue sample containing
a vascular geometry extracted from the rat brain cortex~\citep{motti1986,secomb2000theoretical}.
Inlets and outlets are annotated in the data set.
For the inlets, velocity estimates based on the vessel radius are given in~\citep{secomb2000theoretical}, and herein enforced as Neumann boundary conditions.
The vessel radii are in the range of \SIrange{2}{4.5}{\micro\m}. We use the identical setup as described in \citep{Koch2019a}.
Dirichlet boundary conditions enforce $p_{\soned,\text{out}} = \SI{1.025e5}{\pascal}$ at the outlets.
The extra-vascular domain $\Omega^\text{ex}$ is given by
a rectangular box, $\SI{200}{\um} \times \SI{210}{\um} \times \SI{190}{\um}$.
All boundaries $\partial\Omega \setminus \Gamma$ are considered symmetry boundaries,
$\grad p_\sthreed \cdot \boldsymbol{n} = 0$ on $\partial\Omega$.
The full network geometry and the embedding tissue cube is shown in \cref{fig:bifurcations} (right).

A reference solution is computed using the \textsc{ps} method with
$h_\Omega \in [\SI{0.3}{\micro\m}, \SI{10}{\micro\m}]$ and $h_\Lambda = \SI{1.0}{\micro\m}$
for which we verified grid independence.
The unstructured tetrahedron mesh $\Omega_h$ is locally refined around the vessel
and has $1.1$ Mio. cells, see \cref{fig:bifurcations} (right).
The discrete source terms $q_{K_\Lambda}$ (as defined in \cref{eq:source_discrete})
are computed for $h_\Lambda = \SI{1.0}{\micro\m}$ and
for different $h_\Omega$ using the methods \textsc{css}, and \textsc{ds}
(with kernel radius $\varrho/R_i = 5$, cf.~\citep{Koch2019a}). We start
from $h_\Omega = \SI{20}{\micro\m}$ and refine the grid $\Omega_h^\text{ex}$ (structured Cartesian grid) uniformly.
The total mass flux exchanged between tissue and vessels is computed as
\begin{equation}
  q^{\leftrightarrow}_\Sigma := \frac{1}{2}\sum\limits_{K_\Lambda \in \Lambda_h} \vert q_{K_\Lambda} \vert.
\end{equation}
Moreover, we compute relative differences of the source terms between the implicit interface method solutions
and the reference, i.e. $\lVert \vec{q}^{\textsc{m}} - \vec{q}^\textsc{ps} \rVert_2 / \lVert \vec{q}^\textsc{ps} \rVert_2$,
$M \in \{\textsc{css}, \textsc{ds}\}$, where $\vec{q}$ are vectors with entries $q_{K_\Lambda}$.
To further distinguish errors around bifurcation, we define a set of bifurcation region cells
containing all cells $K_\Lambda$ whose centroid is closer than $\SI{10}{\micro\m}$ to a junction point.

\begin{figure}[htb!]
  \includegraphics[width=1.0\textwidth]{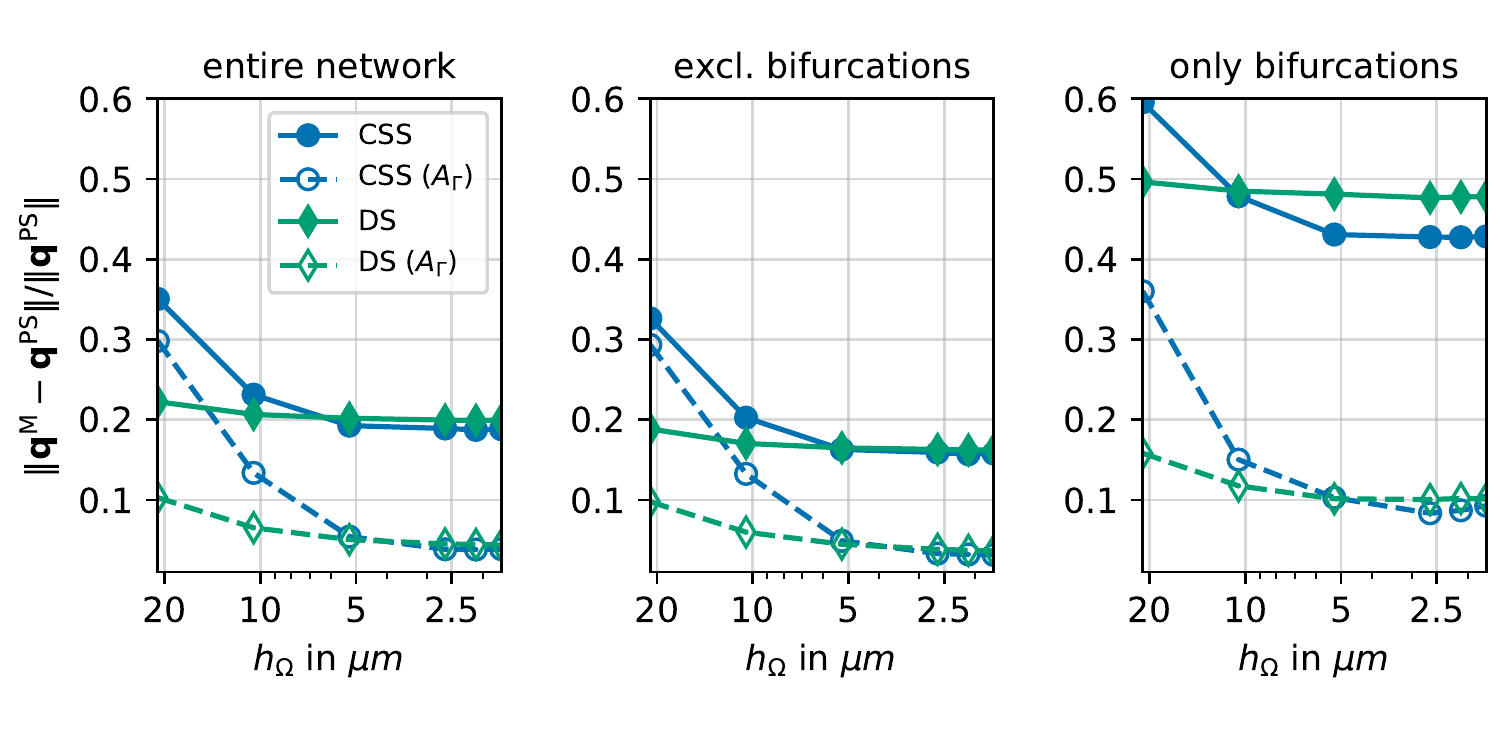}%
  \caption{\textbf{Local effect of (poor) bifurcation geometry approximations.}
  Differences in cell-local source terms between the implicit interface methods \textsc{css} and \textsc{ds}
  and a reference solution obtained with a fine grid and the explicit interface \textsc{ps} method of this work.
  The vectors $\vec{q}$ are vectors of integrated cell-local source terms $q_{K_\Lambda}$ (units of \si{\kg\per\s}).
  In the cases marked with ($A_\Gamma$) the local vessel surface area for each network grid cell $K_\Lambda$ is
  adjusted such that it matches exactly that of the explicitly meshed interface.}
  \label{fig:error_at_bifus}
\end{figure}
Differences in source terms are reported in~\cref{fig:error_at_bifus} (solid lines).
The difference initially decreases with grid refinement but quickly plateaus for
resolutions below $\SI{5}{\micro\m}$. When only looking at the bifurcations regions
(right-most graph in \cref{fig:error_at_bifus} (solid lines)), it is evident that
this difference seems to be concentrated around bifurcations. The reason for this
becomes evident in \cref{fig:bifurcations} which shows differences in the interfacial
area of each network cell $K_\Lambda$ which linearly scales the source term $q_{K_\Lambda}$.
The approximation of each vessel branch with cylindrical segments in the implicit interface methods
introduces local errors in the estimated interfacial area (here in comparison with
the explicitly meshes surface used in the $\textsc{ps}$ method as described in \cref{sec:parametr}).

In a second experiment, we therefore correct the source terms by the area ratio
such that the interfacial area matches the area of the explicit scheme. The results
are shown in ~\cref{fig:error_at_bifus} (dashed lines). The difference at bifurcations
is significantly reduced (from $\approx\SI{45}{\percent}$ to \SI{10}{\percent}).
It also reduces the difference in the rest of the domain and the norm
$\lVert \vec{q}^{\textsc{m}} - \vec{q}^\textsc{ps} \rVert_2 / \lVert \vec{q}^\textsc{ps} \rVert_2$ is reduced
to less than $\SI{3}{\percent}$. We conclude that in case some better information about
the interfacial area is available, the accuracy of implicit interface methods can be improved
by simply accounting for the mismatch in the interfacial area.

\begin{figure}[htb!]
  \includegraphics[width=1.0\textwidth]{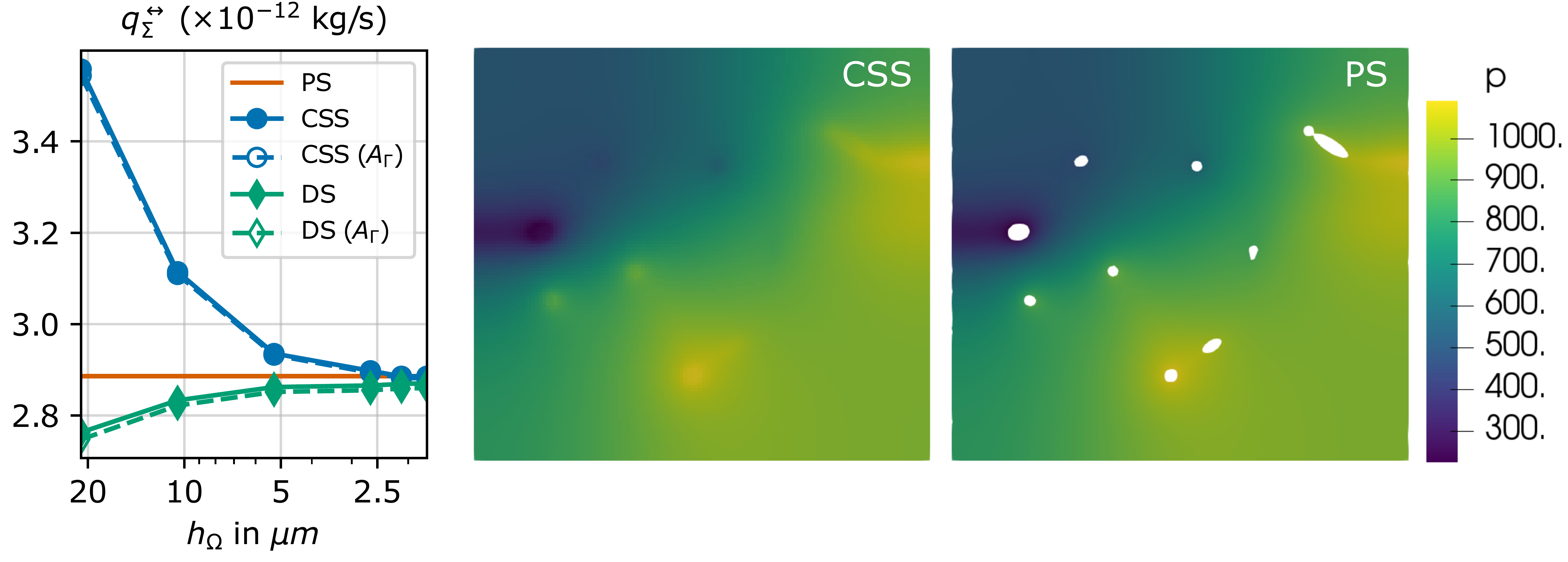}%
  \caption{\textbf{Comparison of total fluid exchange and pressure field.}
  Left, total fluid exchange between network and bulk for the implicit interface methods
  \textsc{css} and \textsc{ds} for different grid resolutions and the reference \textsc{ps} method solution on a fine grid.
  The difference in $q_\Sigma^\leftrightarrow$ between the implicit interface methods and the explicit
  interface method is less than \SI{1}{\percent} on the finest grid and \SI{4}{\percent} (\textsc{ds}) to \SI{23}{\percent} (\textsc{css})
  on the coarsest grid. The correction of the surface area ($A_\Gamma$)
  has no significant influence on the approximation of $q_\Sigma^\leftrightarrow$.
  Right, comparison of pressure field for the \textsc{css} method ($h_\Omega = \SI{1.75}{\micro\m}$)
  and the \textsc{ps} method ($h_\Omega \in [\SI{0.3}{\micro\m}, \SI{10}{\micro\m}]$)
  on a slice through the middle of the domain (without interface area correction).}
  \label{fig:error_at_bifus_global}
\end{figure}
However, when looking at the total fluid exchange $q^{\leftrightarrow}_\Sigma$
and the pressure field in \cref{fig:error_at_bifus_global}, this does not even seem
to be necessary to reproduce accurate results. A possible reason is found in
\cref{fig:bifurcations} for the circled bifurcation $A$ and the kink $C$. Often,
an overestimation of the interfacial area on one side of the bifurcation is
balanced with the underestimation of the interfacial area in a connected vessel branch.
Also it can be seen that these effects are very localized around such features.
Therefore it seems that the pressure field in some distance or the global flux exchange
in a larger tissue volume is hardly affected by these local perturbations of the interfacial area.
Both tested implicit interface methods \textsc{css} and \textsc{ds} show a difference
in $q^{\leftrightarrow}_\Sigma$ of less than \SI{1}{\percent} (for the finest grid) to the explicit interface
method \textsc{ps}. A visual comparison of the pressure maps on a slice obtained
with the \textsc{css} and the \textsc{ps} methods shows an excellent agreement.

In conclusion for the example of fluid tissue perfusion (a linear and stationary
elliptic mixed-dimensional equation system), the new explicit interface method helped
to analyze the suitability of several fundamental assumptions and simplifications
in the derivation of implicit interface methods. Our results show that in the
chosen numerical example with a realistic vessel network and parameters, the
tested implicit interface methods provide very good approximations of the solution
and the assumptions going in the derivation are justified.
In \citep{Koch2019a}, different implicit interface methods have been compared to each other
but no reference model was available. The current results show that in comparison
with an impartial reference solution, the implicit interface methods perform similar in the limit
of fine grids (cf. \cref{fig:error_at_bifus}). This suggests that differences among
the tested implicit interface methods are less relevant
than the modeling error introduced by some common underlying assumptions.
The results also support the finding of \citep{Koch2019a} that the \textsc{ds}
method accurately approximates
(difference in $q^{\leftrightarrow}_\Sigma$ of \SI{4}{\percent} to \textsc{ps}
reference for a resolution of \SI{20}{\micro\m})
the exchange fluid fluxes even for relatively coarse grids.

\subsection{Root water uptake}
\label{sec:root}

In the following application scenario, we compute root water uptake with small root system architecture
obtained from MRI measurements. The scenario is similar to benchmark scenario C1.2 presented in \citep{Schnepf2019benchmark}.
However, instead of a transient problem, we solve a stationary problem
for various root collar pressures enforced as Dirichlet boundary
conditions at the root collar.

The nonlinear mixed-dimensional equation system describing root water uptake
has been introduced in \cref{sec:projintro}. The particularity of this system in contrast to
the previous example of fluid tissue perfusion is that the soil embedding the
root systems is unsaturated leading to complex fluid mechanics involving two fluid phases in porous media.
As roots take up water, the soil dries out in their immediate surrounding (the ratio of air to water
content in the pore space increases). However, the soil's hydraulic conductivity
decreases nonlinearly and overproportionally with water content and likewise water pressure decreases
nonlinearly and overproportionally with water content due to capillary forces. This results in large
pressure gradients at the root soil interface. In soils with low water content
pressure gradients can be several orders of magnitudes larger than in the linear single phase
flow regime in rat brain tissue.

Due to the nonlinearity, the \textsc{ds} method cannot straight-forwardly applied as it relies
on a local interface reconstruction techniques which assumes a linear elliptic PDE. In the following,
we therefore only compare the \textsc{css} method with the \textsc{ps} method.
An extension of the \textsc{ds} method for the case of root water uptake
and a comparison with the \textsc{ps} method is presented in~\citep{Koch2021}.

\begin{figure}[!htb]
	\centering
	\includegraphics[width=0.79\textwidth]{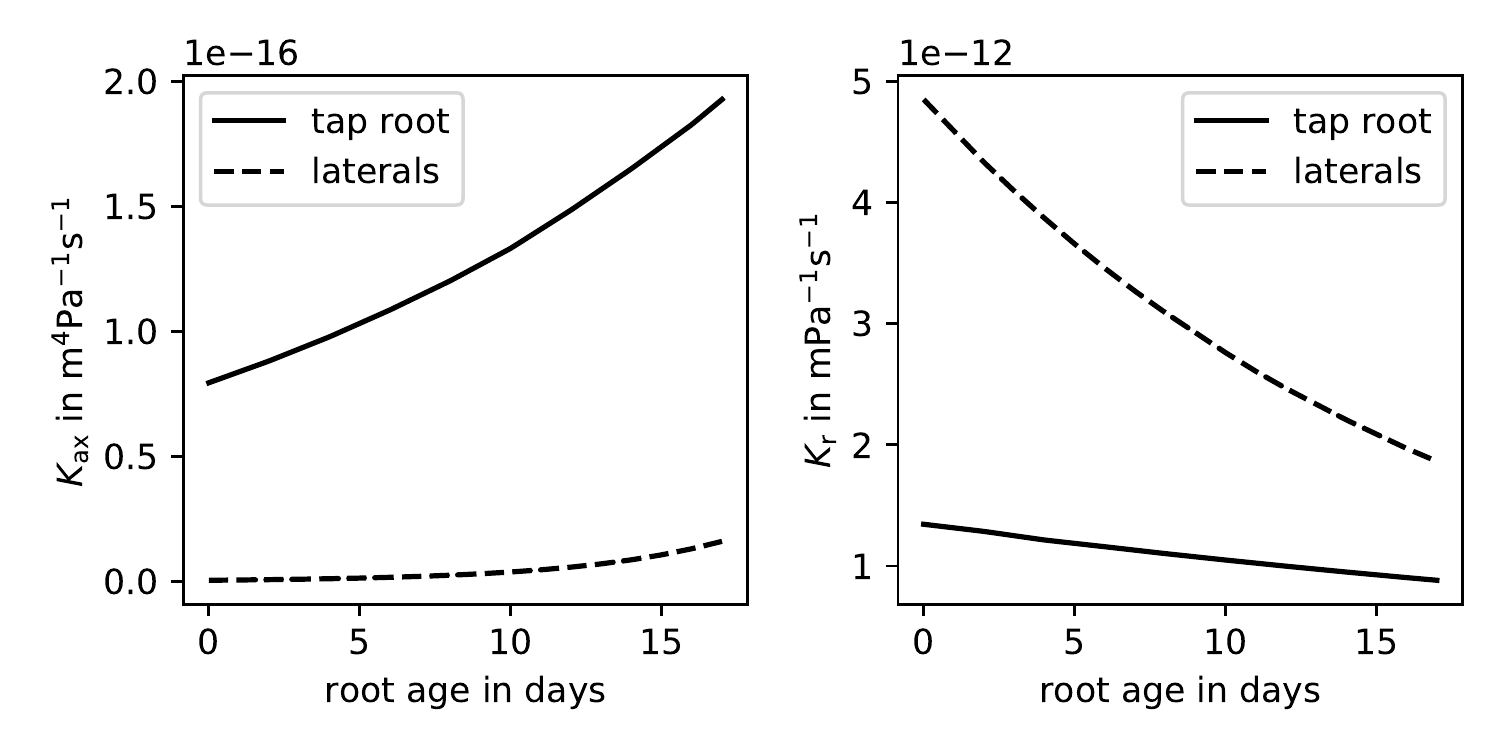}
	\includegraphics[width=0.20\textwidth]{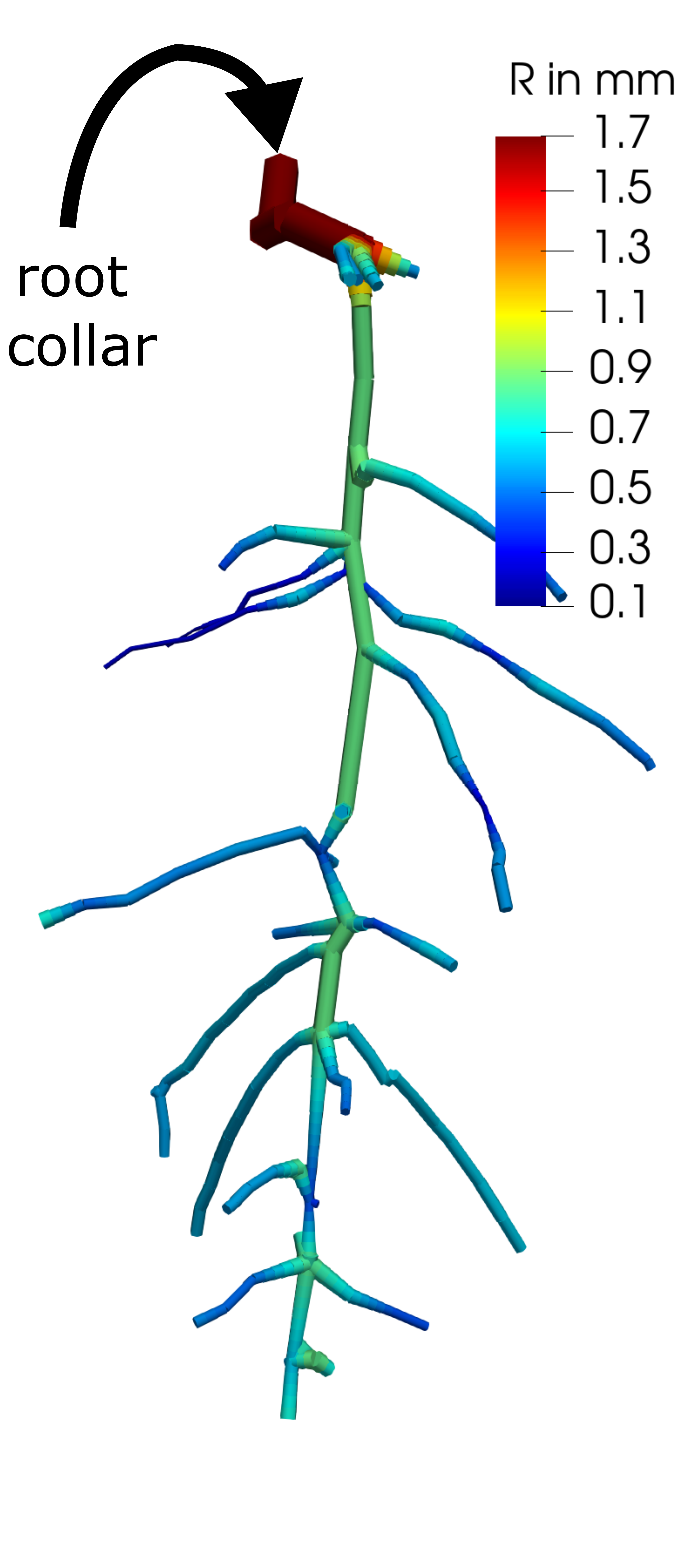}
	\caption{\textbf{Root conductivities and radius for a lupin root system.}
	Left and middle, age-dependent hydraulic root conductivities from~\citep{Schnepf2019benchmark}.
	Right, 8-day-old lupin root system reconstructed from MRI data (courtesy of M. Landl, FZ Jülich).
	Grid data available from \url{https://doi.org/10.18419/darus-471}.
	The root segment radius is visualized to scale. The rooting depth is about \SI{10}{\centi\m}.
	Figure adapted from \citep{Koch2020PhDthesis}.}
	\label{fig:root-conductivities-geometry}
\end{figure}

\begin{table}
  \centering
  \caption{Parameter values for root water uptake example}
  \label{tab:root}
  \begin{tabular}{l l l}
    \toprule
    parameter & value & unit \\\midrule
    $K$ & \num{5.89912e-13} & \si{\square\m} \\
    $\theta_r$ & \num{0.08} & - \\
    $\theta_s$ & \num{0.43} & - \\
    $\alpha_\textsc{vg}$ & \num{4.077e-4} & \si{\per\pascal} \\
    $n_\textsc{vg}$ & \num{1.6} & - \\
    $l_\textsc{vg}$ & \num{0.5} & - \\
    $K_\text{ax}$ & varying, see \cref{fig:root-conductivities-geometry} & \si{\m\tothe{4}\per\pascal\per\s} \\
    $K_\text{r}$ & varying, see \cref{fig:root-conductivities-geometry} & \si{\m\per\pascal\per\s} \\
    \bottomrule
  \end{tabular}
\end{table}

The relationships between hydraulic conductivity and water saturation
(the ratio of water volume to air volume in the pore space) and water pressure
and water saturation can be described by the Van Genuchten-Mualem model~\citep{mualem1976,van1980closed,helmig1997multiphase}.
Parameters for the Van Genuchten-Mualem model are given in \cref{tab:root}, corresponding to a loamy soil, cf. \citep{Schnepf2019benchmark}.
The axial and radial root conductivities vary along the roots dependent on the root age.
These root conductivity values are plotted in \cref{fig:root-conductivities-geometry}.
For tabularized values, we refer to \citep{Schnepf2019benchmark}.
The root system shown in \cref{fig:root-conductivities-geometry} is embedded in a box-shaped domain
with dimension $8\times8\times15$~\si{\centi\m}. The top of the box intersects with the root collar
at $x_3 = \SI{0}{\centi\m}$. The bottom of the domain is located at $x_3 = \SI{-15}{\centi\m}$. We
prescribe a water saturation of $S_w = 0.4$ (corresponding to $p_\sthreed = \SI{0.78e5}{\pascal}$) at all
sides except for the top boundary where we enforce a zero-flow Neumann boundary condition.
In the root domain, we prescribe no-flow boundary conditions at root tips and a fixed
pressure $p_{\soned,c}$ at the root collar. We solve the same scenario for
$p_{\soned,c} = \lbrace \num{0.0}, \num{-0.5e5}, \num{-1.0e5}, \num{-2.5e5}, \num{-5.0e5} \rbrace \si{\pascal}$.
With decreasing root pressure,
the flow rate of water leaving the domain at the root collar (transpiration rate) increases
and the root-soil interface dries out. Dry soil (low water saturation) corresponds to
a strong decrease of the local hydraulic conductivity and low soil water pressures.

\begin{figure}[htb!]
  \includegraphics[width=1.0\textwidth]{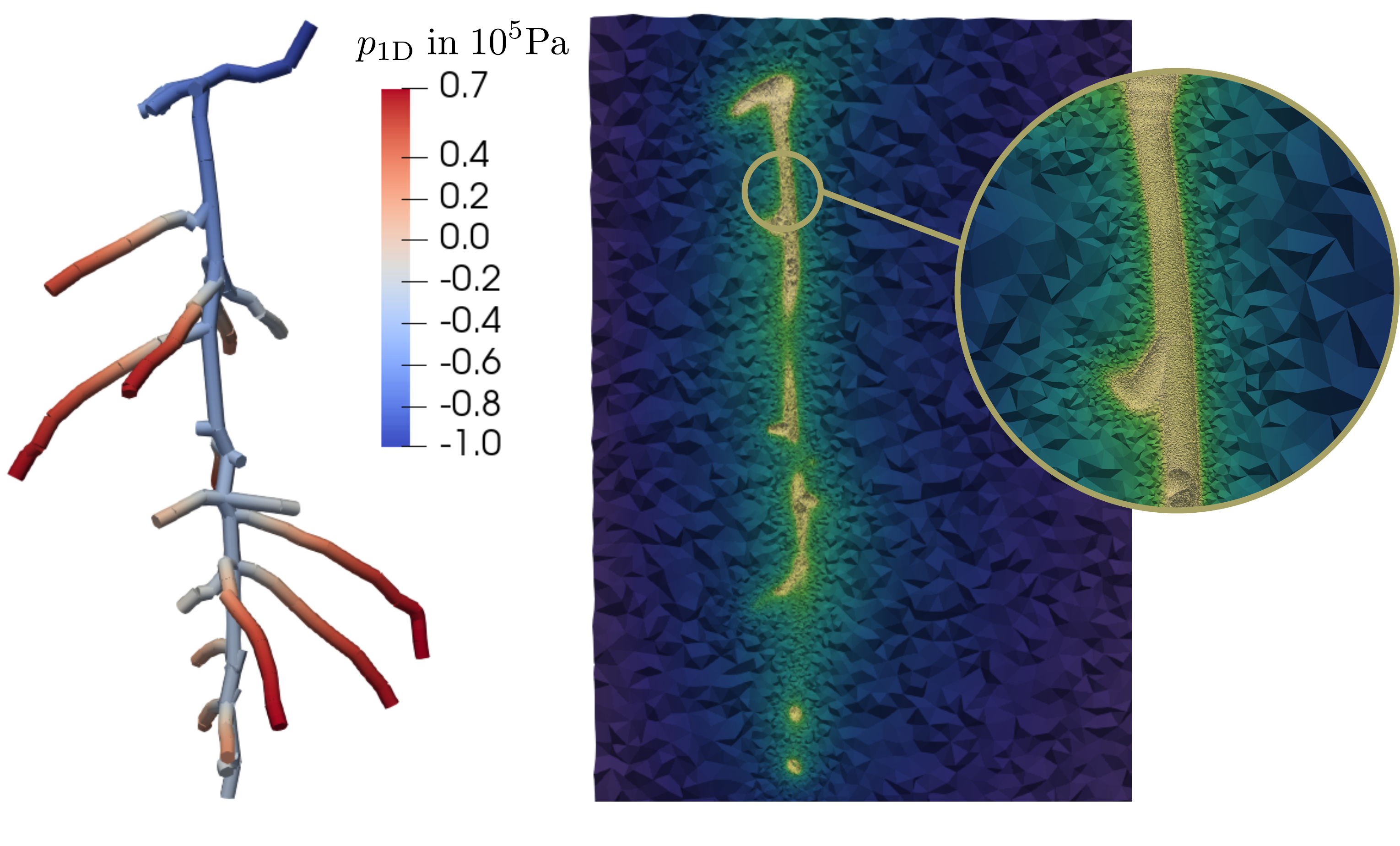}%
  \caption{\textbf{Visualization of root pressure and water saturation for root water uptake scenario.} Stationary solution of
  root pressure (left) and soil water saturation (right) for the case $p_{\soned,c} = \SI{-1e5}{\pascal}$.
  The soil grid is locally refined and resolves the root soil interface (\textsc{ps} method as described in this work).
  Soil water saturation is reduced in the neighborhood of the roots as the root system drains water
  from the soil.}
  \label{fig:rootsoil3d}
\end{figure}

A simulation result for $p_{\soned,c} = \SI{-1e5}{\pascal}$ and the method \textsc{ps}
is shown in \cref{fig:rootsoil3d}. Due to the age dependency of the root hydraulic
conductivities the younger lateral branches have high radial conductivities (enhancing uptake rates)
and relatively low axial conductivities leading to large pressure gradients in such branches.
The opposite is observed in the tap root which is axially conductive but less conductive
in radial direction (reduced uptake rate).
A close-up shows the locally refined grid necessary to accurately resolve the root-soil interface.

\begin{figure}[htb!]
  \includegraphics[width=1.0\textwidth]{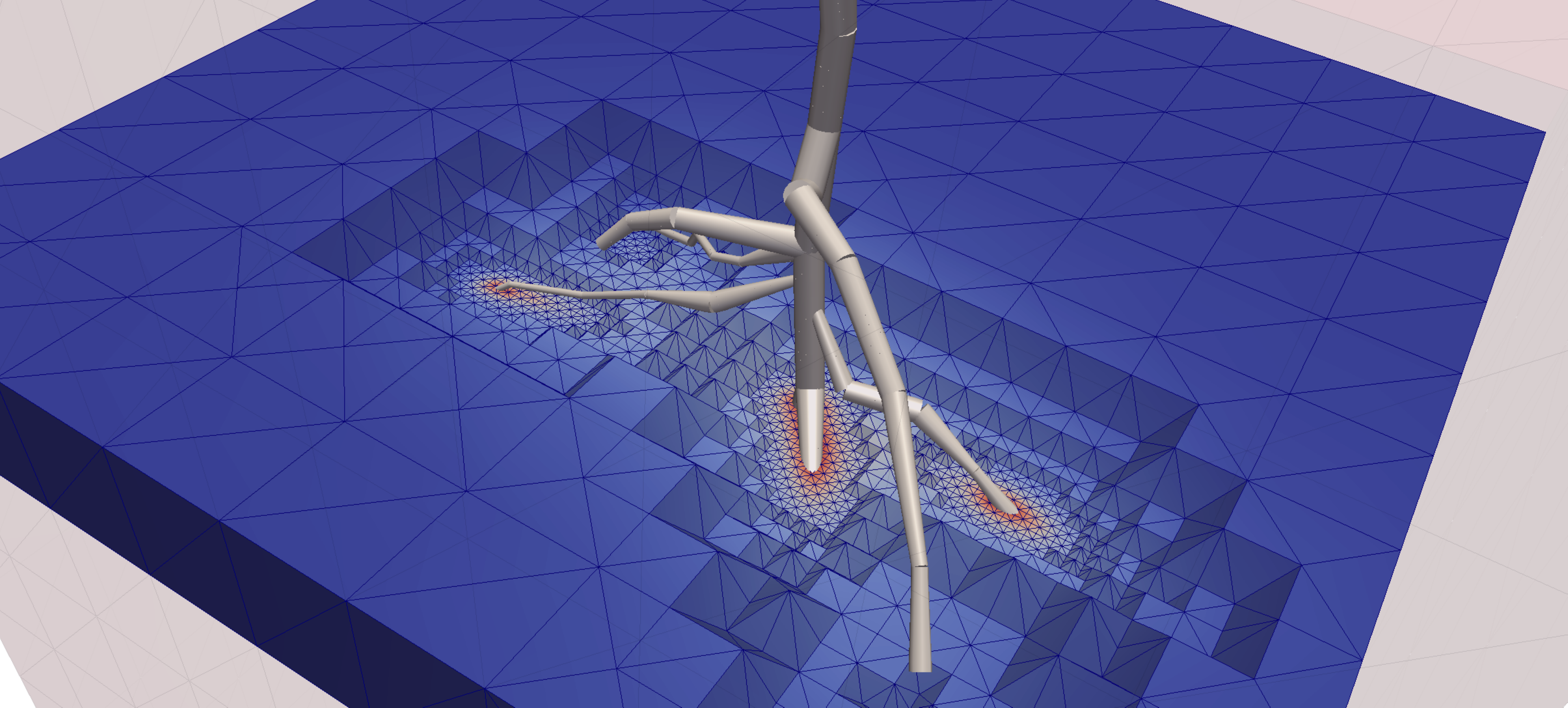}%
  \caption{\textbf{Local grid refinement to resolve pressure gradients at root-soil interface.}
  Computational grid for the implicit interface method \textsc{css}. Color
  shows water saturation, blue corresponds to high saturation ($0.4$) and
  red to low saturation ($0.2$). The soil dries out locally around the roots leading to
  large and strongly localized pressure gradients at the root-soil interface.
  To resolve these gradients the grid $\Omega^\text{ex}_h$ has to be locally refined.
  Figure reprinted from~\citep{Koch2020PhDthesis}.}
  \label{fig:localref}
\end{figure}
The coarsest possible grid resolution of the \textsc{ps} method is limited by the fact that
the root-soil interface needs to be resolved by the mesh. However, as it will become evident in the following
results the fact that pressure gradients become very large in a small neighborhood around
the roots, also requires the \textsc{css} method to use locally refined grids, see \cref{fig:localref}.
To describe the discretization length around the interface,
we introduce $\overline{h}_{10}$ as the average cell diameter
of smallest ten percent of the cells in the soil domain.
As a global measure of how accurate
the source terms $q$ are approximated, we compute the transpiration rate at the root collar. Due
to mass conservation, the transpiration rate can be computed as $r_T = \int_\Lambda q(s) \,\text{d}s$.

\begin{figure}[htb!]
  \includegraphics[width=1.0\textwidth]{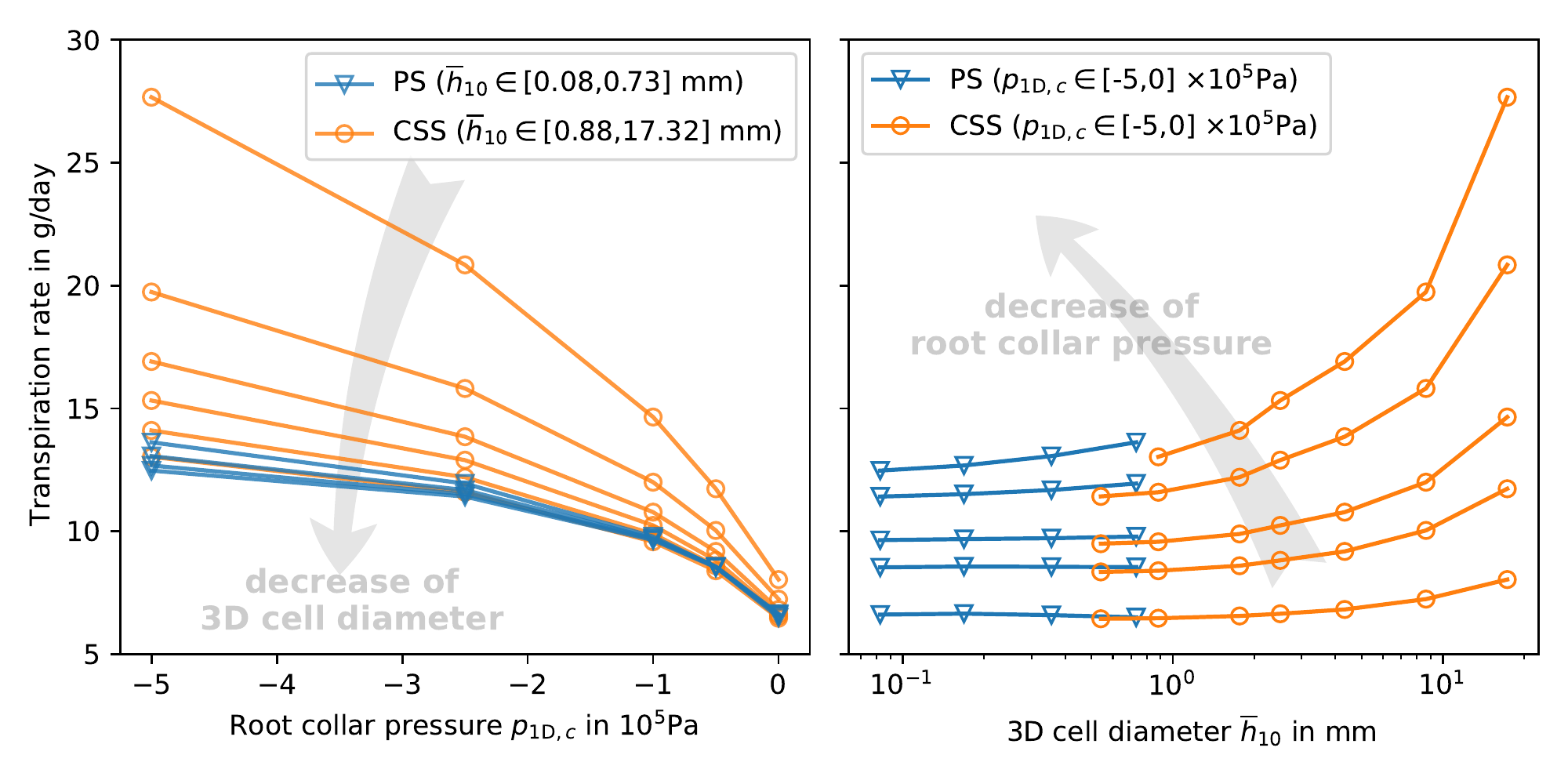}%
  \caption{\textbf{Grid convergence of transpiration rate.} Transpiration rates at the root collar
  for different grid resolutions and the methods \textsc{ps} (explicit interface) and \textsc{css} (implicit interface).
  Left, transpiration rates over
  root collar pressure for different grid resolutions $\overline{h}_{10}^\textsc{ps} = 0.73, 0.36, 0.17$, and $0.08$~\si{\milli\m}
  and $\overline{h}_{10}^\textsc{css} = 17.32, 8.66, 4.33, 2.50, 1.77$ and $0.88$~\si{\milli\m}. Right,
  transpiration rate over grid resolution for different root collar pressures
  $p_{\soned,c} = \num{0.0}, \num{-0.5e5}, \num{-1.0e5}, \num{-2.5e5}$, and \SI{-5.0e5}{\pascal}.
  }
  \label{fig:transpiration}
\end{figure}
To verify the accuracy of the simulation results, we ran simulations
for both the \textsc{ps} method and the \textsc{css} method with different grid resolutions.
\Cref{fig:transpiration} shows the resulting transpiration rates for various grid resolutions and root collar pressures.
For both methods, the transpiration rate decreases with grid refinement. Reasonable grid independence
for the \textsc{ps} method is reached for all cases with the smallest refinement ($\approx 12$Mio. grid cells).
For larger cell diameters as common in root water uptake modelling
with implicit interface methods~\citep{Javaux2008}, the \textsc{css} method significantly
overestimates transpiration rates even for moderately low root collar pressures.
After significant local grid refinement such that soil cell sizes are in
the order of magnitude of the root radius, the \textsc{css} method agrees
reasonably well with the explicit interface method (less than \SI{5}{\percent} difference
in predicted transpiration rate in the worst case: $p_{\soned,c} = \SI{-5.0e5}{\pascal}$).

\begin{figure}[htb!]
  \includegraphics[width=1.0\textwidth]{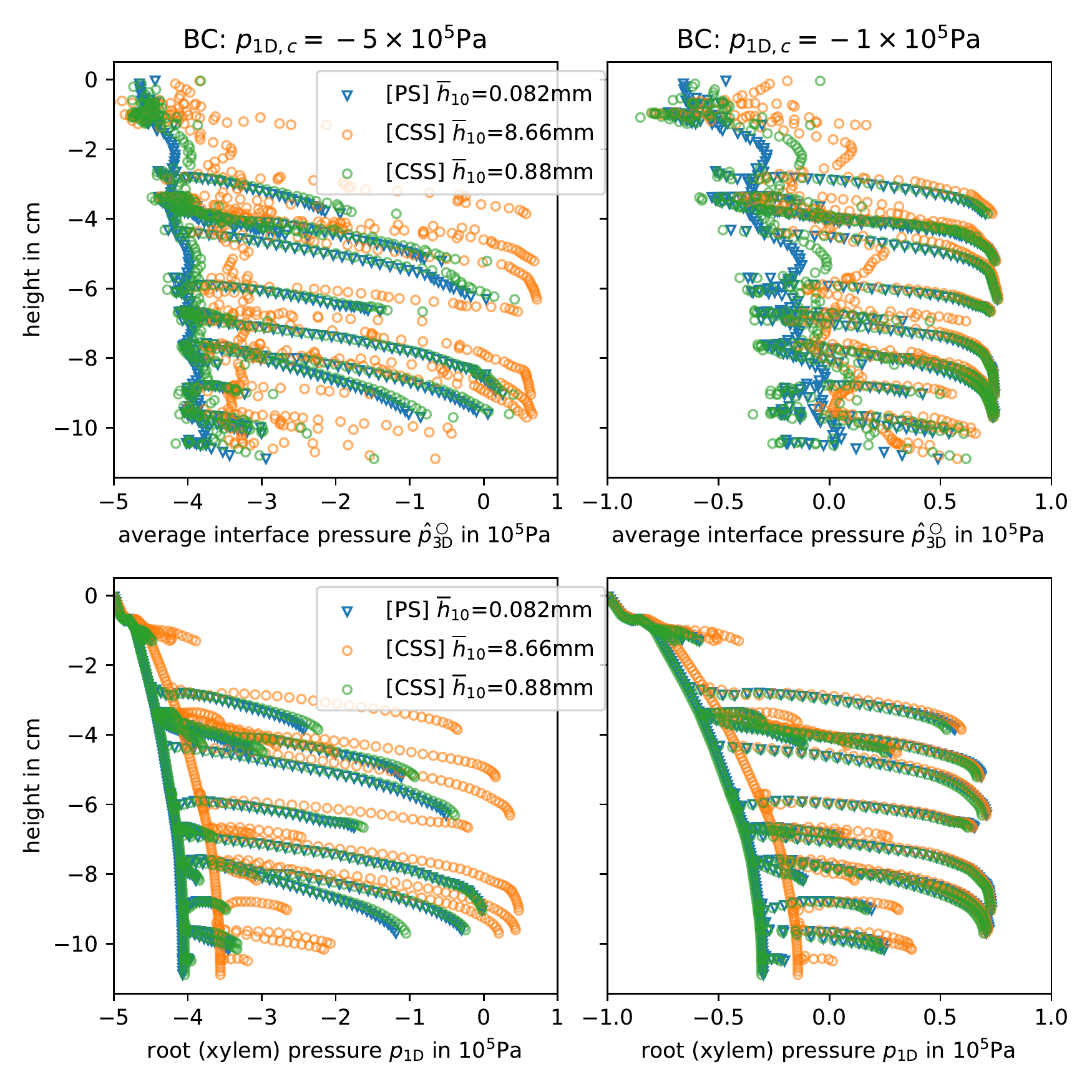}%
  \caption{\textbf{Root pressure and root-soil interface pressure.}
  Top row shows root-soil interface pressure and bottom row shows root pressures for
  every discrete cell $K_\Lambda$. The left column shows
  the scenario $p_{\soned,c} = \SI{-5e5}{\pascal}$,
  the right column shows $p_{\soned,c} = \SI{-1e5}{\pascal}$ prescribed as boundary condition
  at the root collar. All graph compare the solution of the explicit interface \textsc{ps} method
  on a fine grid in comparison with the implicit interface \textsc{css} method for both a fine
  and a coarse grid.}
  \label{fig:reconstr}
\end{figure}

\Cref{fig:reconstr} shows the root pressure ($p_\soned$) and the average root-soil interface pressure
($\hat{p}^\bigcirc_{\sthreed}$) for every cell $K_\Lambda \in \Lambda_h$ and two root collar
pressure boundary conditions $p_{\soned,c} = \SI{-5e5}{\pascal}$ (low) and $p_{\soned,c} = \SI{-1e5}{\pascal}$ (moderate).
Pressures are plotted over the height in the soil discarding the information of the horizontal position.
Since pressure gradients are large in the lateral roots, lateral roots and the tap roots can be clearly
distinguished in the projected plots. In case of moderate root collar pressure both root and interface
pressures are high in lateral roots corresponding to high water saturation and high soil conductivity.
Therefore, pressure gradients are relatively small and are well approximated even with the coarsest grid
($\overline{h}_{10}^\textsc{css} = \SI{8.66}{\milli\m}$). However, in the tap root where pressures are lower,
we observe a significant difference between the methods for coarse grids. This explains the large difference
in transpiration rates for such grids seen in \cref{fig:transpiration}. The result significantly improves
with local grid refinement leading to a good match between \textsc{ps} and \textsc{css} method.
For the low pressure (dry soil) case, a strong mismatch between
the fine grid \textsc{ps} reference and the coarse grid \textsc{css}
can be observed in both lateral roots and the tap root. Again both methods agree reasonably well
when the grid is locally refined. Interestingly the root-soil interface pressure is significantly
higher between branches then around joints. This can be explained with the fact that the total root
density is increased around joints and the local interfacial area and the local water uptake
is higher in such regions.

Although it is not investigated properly in this work, we conclude with a brief comment concerning computational efficiency.
To this end, we note that the grid resolutions $\overline{h}_{10}^\textsc{ps} = 0.73, 0.36, 0.17$, and $0.08$~\si{\milli\m}
correspond to discretizations with $135$k, $264$k, $720$k, $2.3$M degrees of freedom ($755$k, $1.4$M, $3.8$M, $12$M cells),
and the resolutions used for the \textsc{css} method, $\overline{h}_{10}^\textsc{css} = 17.32, 8.66, 4.33, 2.50, 1.77$ and $0.88$~\si{\milli\m},
correspond to discretizations with $46$k, $59$k, $83$k, $170$k, $180$k, $458$k degrees of freedom ($9$k, $11$k, $16$k, $31$k, $1.0$M, $2.7$M cells).
While the solver time scales with the number of degrees of freedom, the assembly time scales with the number of cells.
In all cases, only $600$ degrees of freedom are needed to discretize the root domain accurately enough.
We find the \textsc{css} in our implementation to be less efficient than the \textsc{ps} method (when using the same
amount of degrees of freedom) which can be attributed
to the non-local stencil (stencil increases with refinement) due the average operator to compute $\hat{p}^\bigcirc_{\sthreed}$.
The stencil of the \textsc{ps} method is local in the sense that degrees of freedom are only coupled with degrees of freedom
in the immediate neighborhood. Furthermore, we found that number of Newton iterations to be slightly higher on
average for the \textsc{css} method. Depending on the requirements on accuracy, the \textsc{css} method may be used
with a slightly coarser grid than the \textsc{ps} method, however to obtain a difference of e.g. less than \SI{5}{\percent}
in the transpiration rate, similar grid resolutions are necessary ($\approx 500$k degrees of freedom in the presented example).
Hence, perhaps somewhat surprising, the explicit interface method is not much less efficient than the \textsc{css} method
for root water uptake simulations in dry soil. However, arguably the meshing procedure is more involved for the \textsc{ps} method.

\section{Summary and conclusion}
\label{sec:summary}

Flow and transport problems featuring embedded tubular network systems arise in many
biological and technical applications such as root water and nutrient uptake,
fluid perfusion of vascularized tissues, well modeling in geothermal or petroleum reservoirs,
or heat exchangers. Mixed-dimension methods where the embedded network is reduced to a system
of one-dimensional PDEs coupled with three-dimensional PDEs for the transport in the embedding bulk domain
are efficient methods to simulate flow and transport in such systems.

We introduced a new mixed-dimension method which explicitly resolves the bulk-network interface,
in the bulk mesh, while the network is still described with one-dimensional PDEs. We related
the new explicit interface method to commonly used implicit interface methods.
While resolving the bulk-network interface requires a high effort concerning the generation of computational meshes,
it allows to simulate time-dependent and nonlinear problems using standard discretization
techniques in both subdomains. In contrast, methods with implicit surface descriptions
(which may allow the use of completely structured bulk meshes that do not resolve the interface,)
often require additional model assumptions and assume linear elliptic PDEs.
However, there is a strong need to investigate the accuracy
of efficient implicit interface methods for time-dependent and nonlinear problem,
for instance, for the modeling of tracer perfusion in vascularized tissue, or root water uptake from soil.
We therefore see the presented resolved-interface model as a good candidate for benchmarking
new mixed-dimension methods and as a sound and feasible alternative to comparing with fully three-dimensional models.

For the introduced interface resolving method, we suggested a practical surface description if
only a centerline network and a radius function is given to describe the network domain.
Furthermore, we suggested an efficient integration scheme for the source terms coupling
network and bulk problems.

We used the introduced method in numerical comparisons with implicit interface mixed-dimension methods in two
application cases: (a) The simulation of fluid flow in vascularized tissue with a small network extracted
from a rat brain, modeled by a linear elliptic mixed-dimensional PDE systems. (b) The simulation
of root water uptake from loamy soil with a small lupin root network extracted from MRI images, modeled
by a strongly nonlinear elliptic mixed-dimensional PDE system. Regarding the
numerical investigations conducted with the resolved-interface method in the role of a reference solution,
we summarize the following conclusions.
For the case of fluid perfusion of vascularized tissue,
we found that the error made by neglecting the vessel resistance to bulk flow is insignificant.
Simple cylinder approximations of the vessels may introduce local errors in the interfacial area
affecting the predicted fluid exchange. However, we have found these errors to be insignificant when
looking at the pressure distribution (in particular in a small distance to vessels) and total fluid
exchange in a small region of interest with several vessels. Therefore, implicit interface methods produce
both efficient and accurate results.
It became evident in the root water uptake case that the resolution of local pressure gradients
in the soil becomes the limiting factor when determining grid resolutions that yield accurate results.
Not resolving the length scale of the drop in soil water pressure and soil hydraulic conductivity leads
to large errors in the estimation of transpiration rates even for moderate pressures. This error
by far dominates the geometry-related model errors investigated in the tissue perfusion example.
The result strongly suggests the need for implicit interface methods that can overcome this
problem of grid resolution, for example concepts using local analytical or numerical solutions
in the immediate neighborhood of the roots~\citep{Schroeder2008localsoil,Beudez2013,Mai2019,Koch2021}.

\section*{Acknowledgements}

The author would like to thank Martin Schneider
for proof-reading the initial manuscript.
The research leading to these results has received funding
from the European Union's Horizon 2020 research and innovation programme under the Marie Skłodowska-Curie grant agreement No 801133
and from the German Research Foundation (DFG), within
the Collaborative Research Center on Interface-Driven Multi-Field Processes in Porous Media (SFB 1313, Project No. 327154368).

\section*{Data availability statement}

The source code and an example application
will be made freely available under an open source license
as part of the next version of the
porous medium simulator \dumux \citep{Koch2019cdumux}, \url{https://dumux.org}.

\bibliography{projection}

\begin{thebibliography}{10}
\expandafter\ifx\csname url\endcsname\relax
  \def\url#1{\texttt{#1}}\fi
\expandafter\ifx\csname urlprefix\endcsname\relax\def\urlprefix{URL }\fi
\expandafter\ifx\csname href\endcsname\relax
  \def\href#1#2{#2} \def\path#1{#1}\fi

\bibitem{Reichold2009}
J.~Reichold, M.~Stampanoni, A.~L. Keller, A.~Buck, P.~Jenny, B.~Weber, Vascular
  graph model to simulate the cerebral blood flow in realistic vascular
  networks, Journal of Cerebral Blood Flow \& Metabolism 29~(8) (2009)
  1429--1443.
\newblock \href {http://dx.doi.org/10.1038/jcbfm.2009.58}
  {\path{doi:10.1038/jcbfm.2009.58}}.

\bibitem{Koch2019cdumux}
T.~Koch, D.~Gläser, K.~Weishaupt, S.~Ackermann, M.~Beck, B.~Becker,
  S.~Burbulla, H.~Class, E.~Coltman, S.~Emmert, T.~Fetzer, C.~Grüninger,
  K.~Heck, J.~Hommel, T.~Kurz, M.~Lipp, F.~Mohammadi, S.~Scherrer,
  M.~Schneider, G.~Seitz, L.~Stadler, M.~Utz, F.~Weinhardt, B.~Flemisch,
  Du{M}u$^\text{x}$ 3 -- an open-source simulator for solving flow and
  transport problems in porous media with a focus on model coupling, Computers
  \& Mathematics with Applications 81 (2021) 423--443.
\newblock \href {http://dx.doi.org/10.1016/j.camwa.2020.02.012}
  {\path{doi:10.1016/j.camwa.2020.02.012}}.

\bibitem{Blinder2013}
P.~Blinder, P.~S. Tsai, J.~P. Kaufhold, P.~M. Knutsen, H.~Suhl, D.~Kleinfeld,
  The cortical angiome: an interconnected vascular network with noncolumnar
  patterns of blood flow, Nature Neuroscience 16~(7) (2013) 889--897.
\newblock \href {http://dx.doi.org/10.1038/nn.3426}
  {\path{doi:10.1038/nn.3426}}.

\bibitem{Leitner2014b}
D.~Leitner, F.~Meunier, G.~Bodner, M.~Javaux, A.~Schnepf, Impact of contrasted
  maize root traits at flowering on water stress tolerance -- a simulation
  study, Field Crops Research 165 (2014) 125 -- 137.
\newblock \href {http://dx.doi.org/10.1016/j.fcr.2014.05.009}
  {\path{doi:10.1016/j.fcr.2014.05.009}}.

\bibitem{Hsu1989green}
R.~Hsu, T.~W. Secomb, A {Green's} function method for analysis of oxygen
  delivery to tissue by microvascular networks, Mathematical Biosciences 96~(1)
  (1989) 61--78.
\newblock \href {http://dx.doi.org/10.1016/0025-5564(89)90083-7}
  {\path{doi:10.1016/0025-5564(89)90083-7}}.

\bibitem{dangelo2008}
C.~D'Angelo, A.~Quarteroni, On the coupling of 1d and 3d diffusion-reaction
  equations: Application to tissue perfusion problems, Mathematical Models and
  Methods in Applied Sciences 18~(08) (2008) 1481--1504.
\newblock \href {http://dx.doi.org/10.1142/S0218202508003108}
  {\path{doi:10.1142/S0218202508003108}}.

\bibitem{DAngelo2012}
C.~D'Angelo, Finite element approximation of elliptic problems with dirac
  measure terms in weighted spaces: Applications to one- and three-dimensional
  coupled problems, SIAM Journal on Numerical Analysis 50~(1) (2012) 194--215.
\newblock \href {http://dx.doi.org/10.1137/100813853}
  {\path{doi:10.1137/100813853}}.

\bibitem{cattaneo2014computational}
L.~Cattaneo, P.~Zunino, Computational models for fluid exchange between
  microcirculation and tissue interstitium, Networks \& Heterogeneous Media
  9~(1).
\newblock \href {http://dx.doi.org/10.3934/nhm.2014.9.135}
  {\path{doi:10.3934/nhm.2014.9.135}}.

\bibitem{Gjerde2018}
I.~G. Gjerde, K.~Kumar, J.~M. Nordbotten, B.~Wohlmuth, Splitting method for
  elliptic equations with line sources, ESAIM: Mathematical Modelling and
  Numerical Analysis 53~(5) (2019) 1715--1739.
\newblock \href {http://dx.doi.org/10.1051/m2an/2019027}
  {\path{doi:10.1051/m2an/2019027}}.

\bibitem{koeppl2018}
T.~Köppl, E.~Vidotto, B.~Wohlmuth, P.~Zunino, Mathematical modeling, analysis
  and numerical approximation of second-order elliptic problems with
  inclusions, Mathematical Models and Methods in Applied Sciences 28~(05)
  (2018) 953--978.
\newblock \href {http://dx.doi.org/10.1142/S0218202518500252}
  {\path{doi:10.1142/S0218202518500252}}.

\bibitem{Laurino2019}
F.~Laurino, P.~Zunino, Derivation and analysis of coupled {PDEs} on manifolds
  with high dimensionality gap arising from topological model reduction,
  {ESAIM}: Mathematical Modelling and Numerical Analysis 53~(6) (2019)
  2047--2080.
\newblock \href {http://dx.doi.org/10.1051/m2an/2019042}
  {\path{doi:10.1051/m2an/2019042}}.

\bibitem{Koch2019a}
T.~Koch, M.~Schneider, R.~Helmig, P.~Jenny, Modeling tissue perfusion in terms
  of 1d-3d embedded mixed-dimension coupled problems with distributed sources,
  Journal of Computational Physics 410 (2020) 109370.
\newblock \href {http://dx.doi.org/10.1016/j.jcp.2020.109370}
  {\path{doi:10.1016/j.jcp.2020.109370}}.

\bibitem{Doussan1998}
C.~Doussan, L.~Pages, G.~Vercambre, {Modelling of the Hydraulic Architecture of
  Root Systems: An Integrated Approach to Water Absorption—Model
  Description}, Annals of Botany 81~(2) (1998) 213--223.
\newblock \href {http://dx.doi.org/10.1006/anbo.1997.0540}
  {\path{doi:10.1006/anbo.1997.0540}}.

\bibitem{Schroeder2009grid}
T.~Schröder, L.~Tang, M.~Javaux, J.~Vanderborght, B.~Körfgen, H.~Vereecken, A
  grid refinement approach for a three-dimensional soil-root water transfer
  model, Water Resources Research 45~(10), w10412.
\newblock \href {http://dx.doi.org/10.1029/2009WR007873}
  {\path{doi:10.1029/2009WR007873}}.

\bibitem{Beudez2013}
N.~Beudez, C.~Doussan, G.~Lefeuve-Mesgouez, A.~Mesgouez, Influence of three
  root spatial arrangement on soil water flow and uptake. results from an
  explicit and an equivalent, upscaled, model, Procedia Environmental Sciences
  19 (2013) 37--46, four Decades of Progress in Monitoring and Modeling of
  Processes in the Soil-Plant-Atmosphere System: Applications and Challenges.
\newblock \href {http://dx.doi.org/10.1016/j.proenv.2013.06.005}
  {\path{doi:10.1016/j.proenv.2013.06.005}}.

\bibitem{Mai2019}
T.~H. Mai, A.~Schnepf, H.~Vereecken, J.~Vanderborght, Continuum multiscale
  model of root water and nutrient uptake from soil with explicit consideration
  of the 3d root architecture and the rhizosphere gradients, Plant and Soil
  439~(1) (2019) 273--292.
\newblock \href {http://dx.doi.org/10.1007/s11104-018-3890-4}
  {\path{doi:10.1007/s11104-018-3890-4}}.

\bibitem{Schroeder2008localsoil}
T.~Schröder, M.~Javaux, J.~Vanderborght, B.~Körfgen, H.~Vereecken, Effect of
  local soil hydraulic conductivity drop using a three-dimensional root water
  uptake model 7 (2008) 1089--1098, 3.
\newblock \href {http://dx.doi.org/10.2136/vzj2007.0114}
  {\path{doi:10.2136/vzj2007.0114}}.

\bibitem{Schroeder2009implmicro}
T.~Schröder, M.~Javaux, J.~Vanderborght, B.~Körfgen, H.~Vereecken,
  {Implementation of a Microscopic Soil–Root Hydraulic Conductivity Drop
  Function in a Three-Dimensional Soil–Root Architecture Water Transfer
  Model}, Vadose Zone Journal 8~(3) (2009) 783--792.
\newblock \href {http://dx.doi.org/10.2136/vzj2008.0116}
  {\path{doi:10.2136/vzj2008.0116}}.

\bibitem{koeppl2016}
T.~Köppl, E.~Vidotto, B.~Wohlmuth, A local error estimate for the poisson
  equation with a line source term, in: B.~Karas{\"o}zen, M.~Manguo{\u{g}}lu,
  M.~Tezer-Sezgin, S.~G{\"o}ktepe, {\"O}.~U{\u{g}}ur (Eds.), Numerical
  Mathematics and Advanced Applications ENUMATH 2015, Springer International
  Publishing, Cham, 2016, pp. 421--429.

\bibitem{Koch2019bwell}
T.~Koch, R.~Helmig, M.~Schneider, A new and consistent well model for one-phase
  flow in anisotropic porous media using a distributed source model, Journal of
  Computational Physics 410 (2020) 109369.
\newblock \href {http://dx.doi.org/10.1016/j.jcp.2020.109369}
  {\path{doi:10.1016/j.jcp.2020.109369}}.

\bibitem{Daly2018}
K.~R. Daly, S.~R. Tracy, N.~M. Crout, S.~Mairhofer, T.~P. Pridmore, S.~J.
  Mooney, T.~Roose, Quantification of root water uptake in soil using x-ray
  computed tomography and image-based modelling, Plant, Cell \& Environment
  41~(1) (2018) 121--133.
\newblock \href {http://dx.doi.org/10.1111/pce.12983}
  {\path{doi:10.1111/pce.12983}}.

\bibitem{Schnepf2019benchmark}
A.~Schnepf, C.~K. Black, V.~Couvreur, B.~M. Delory, C.~Doussan, A.~Koch,
  T.~Koch, M.~Javaux, M.~Landl, D.~Leitner, G.~Lobet, T.~H. Mai, F.~Meunier,
  L.~Petrich, J.~A. Postma, E.~Priesack, V.~Schmidt, J.~Vanderborght,
  H.~Vereecken, M.~Weber, Call for participation: Collaborative benchmarking of
  functional-structural root architecture models. the case of root water
  uptake, Frontiers in Plant Science 11 (2020) 316.
\newblock \href {http://dx.doi.org/10.3389/fpls.2020.00316}
  {\path{doi:10.3389/fpls.2020.00316}}.

\bibitem{Koch2020PhDthesis}
T.~Koch, Mixed-dimension models for flow and transport processes in porous
  media with embedded tubular network systems, Ph.D. thesis, University of
  Stuttgart (2020).
\newblock \href {http://dx.doi.org/10.18419/opus-10975}
  {\path{doi:10.18419/opus-10975}}.

\bibitem{Javaux2008}
M.~Javaux, T.~Schröder, J.~Vanderborght, H.~Vereecken, {Use of a
  Three-Dimensional Detailed Modeling Approach for Predicting Root Water
  Uptake}, Vadose Zone Journal 7~(3) (2008) 1079.
\newblock \href {http://dx.doi.org/10.2136/vzj2007.0115}
  {\path{doi:10.2136/vzj2007.0115}}.

\bibitem{Roose2008}
T.~Roose, A.~Schnepf, Mathematical models of plant{\textendash}soil
  interaction, Philosophical Transactions of the Royal Society of London A:
  Mathematical, Physical and Engineering Sciences 366~(1885) (2008) 4597--4611.
\newblock \href {http://dx.doi.org/10.1098/rsta.2008.0198}
  {\path{doi:10.1098/rsta.2008.0198}}.

\bibitem{Dunbabin2013}
V.~M. Dunbabin, J.~a. Postma, A.~Schnepf, L.~Pag{\`{e}}s, M.~Javaux, L.~Wu,
  D.~Leitner, Y.~L. Chen, Z.~Rengel, A.~J. Diggle, {Modelling root-soil
  interactions using three-dimensional models of root growth, architecture and
  function}, Plant and Soil 372 (2013) 93--124.
\newblock \href {http://dx.doi.org/10.1007/s11104-013-1769-y}
  {\path{doi:10.1007/s11104-013-1769-y}}.

\bibitem{Koch2018a}
T.~Koch, K.~Heck, N.~Schröder, H.~Class, R.~Helmig, A new simulation framework
  for soil-root interaction, evaporation, root growth, and solute transport,
  Vadose Zone Journal 17.
\newblock \href {http://dx.doi.org/10.2136/vzj2017.12.0210}
  {\path{doi:10.2136/vzj2017.12.0210}}.

\bibitem{mualem1976}
Y.~Mualem, A new model for predicting the hydraulic conductivity of unsaturated
  porous media, Water Resources Research 12~(3) (1976) 513--522.
\newblock \href {http://dx.doi.org/10.1029/WR012i003p00513}
  {\path{doi:10.1029/WR012i003p00513}}.

\bibitem{van1980closed}
M.~T. Van~Genuchten, A closed-form equation for predicting the hydraulic
  conductivity of unsaturated soils, Soil science society of America journal
  44~(5) (1980) 892--898.
\newblock \href {http://dx.doi.org/10.2136/sssaj1980.03615995004400050002x}
  {\path{doi:10.2136/sssaj1980.03615995004400050002x}}.

\bibitem{CGAL2019}
{The CGAL Project},
  \href{https://doc.cgal.org/4.14/Manual/packages.html}{{CGAL} User and
  Reference Manual}, {4.14} Edition, {CGAL Editorial Board}, 2019.
\newline\urlprefix\url{https://doc.cgal.org/4.14/Manual/packages.html}

\bibitem{Boissonnat2005}
J.-D. Boissonnat, S.~Oudot, Provably good sampling and meshing of surfaces,
  Graphical Models 67~(5) (2005) 405--451.
\newblock \href {http://dx.doi.org/10.1016/j.gmod.2005.01.004}
  {\path{doi:10.1016/j.gmod.2005.01.004}}.

\bibitem{Lindquist1996}
W.~B. Lindquist, S.-M. Lee, D.~A. Coker, K.~W. Jones, P.~Spanne, Medial axis
  analysis of void structure in three-dimensional tomographic images of porous
  media, Journal of Geophysical Research: Solid Earth 101~(B4) (1996)
  8297--8310.
\newblock \href {http://dx.doi.org/10.1029/95JB03039}
  {\path{doi:10.1029/95JB03039}}.

\bibitem{Antiga2008}
L.~Antiga, M.~Piccinelli, L.~Botti, B.~Ene-Iordache, A.~Remuzzi, D.~A.
  Steinman, An image-based modeling framework for patient-specific
  computational hemodynamics, Medical {\&} Biological Engineering {\&}
  Computing 46~(11) (2008) 1097--1112.
\newblock \href {http://dx.doi.org/10.1007/s11517-008-0420-1}
  {\path{doi:10.1007/s11517-008-0420-1}}.

\bibitem{smoothmin2013}
I.~Quilez, \href{https://www.iquilezles.org/www/articles/smin/smin.htm}{Smooth
  minimum function},
  \url{https://www.iquilezles.org/www/articles/smin/smin.htm}, last accessed 21
  September 2019.
\newline\urlprefix\url{https://www.iquilezles.org/www/articles/smin/smin.htm}

\bibitem{d2007multiscale}
C.~D'Angelo, Multiscale modelling of metabolism and transport phenomena in
  living tissues, Bibliotheque de l'EPFL, Lausanne.

\bibitem{Hackbusch1989}
W.~Hackbusch, On first and second order box schemes, Computing 41~(4) (1989)
  277--296.
\newblock \href {http://dx.doi.org/10.1007/bf02241218}
  {\path{doi:10.1007/bf02241218}}.

\bibitem{HuberHelmig1999}
R.~Huber, R.~Helmig, Multiphase flow in heterogeneous porous media: A classical
  finite element method versus an implicit pressure-explicit saturation-based
  mixed finite element-finite volume approach, International Journal for
  Numerical Methods in Fluids 29~(8) (1999) 899--920.
\newblock \href
  {http://dx.doi.org/10.1002/(SICI)1097-0363(19990430)29:8<899::AID-FLD715>3.0.CO;2-W}
  {\path{doi:10.1002/(SICI)1097-0363(19990430)29:8<899::AID-FLD715>3.0.CO;2-W}}.

\bibitem{Schneider2018}
M.~Schneider, D.~Gl\"aser, B.~Flemisch, R.~Helmig, Comparison of finite-volume
  schemes for diffusion problems, Oil Gas Sci. Technol. - Rev. IFP Energies
  nouvelles 73 (2018) 82.
\newblock \href {http://dx.doi.org/10.2516/ogst/2018064}
  {\path{doi:10.2516/ogst/2018064}}.

\bibitem{foamgrid}
O.~Sander, T.~Koch, N.~Schröder, B.~Flemisch, The dune foamgrid implementation
  for surface and network grids, Archive of Numerical Software 5~(1) (2017)
  217--244.
\newblock \href {http://dx.doi.org/10.11588/ans.2017.1.28490}
  {\path{doi:10.11588/ans.2017.1.28490}}.

\bibitem{levick1991}
J.~Levick, Capillary filtration-absorption balance reconsidered in light of
  dynamic extravascular factors, Experimental Physiology 76~(6) (1991)
  825--857.
\newblock \href {http://dx.doi.org/10.1113/expphysiol.1991.sp003549}
  {\path{doi:10.1113/expphysiol.1991.sp003549}}.

\bibitem{motti1986}
E.~D.~F. Motti, H.-G. Imhof, M.~G. Yaşargil, The terminal vascular bed in the
  superficial cortex of the rat, Journal of Neurosurgery 65~(6) (1986)
  834--846.
\newblock \href {http://dx.doi.org/10.3171/jns.1986.65.6.0834}
  {\path{doi:10.3171/jns.1986.65.6.0834}}.

\bibitem{secomb2000theoretical}
T.~W. Secomb, R.~Hsu, N.~Beamer, B.~Coull, Theoretical simulation of oxygen
  transport to brain by networks of microvessels: Effects of oxygen supply and
  demand on tissue hypoxia, Microcirculation 7~(4) (2000) 237--247.
\newblock \href {http://dx.doi.org/10.1111/j.1549-8719.2000.tb00124.x}
  {\path{doi:10.1111/j.1549-8719.2000.tb00124.x}}.

\bibitem{Koch2021}
T.~Koch, H.~Wu, M.~Schneider, Nonlinear mixed-dimension model for embedded
  tubular networks with application to root water uptake (2021).
\newblock \href {http://arxiv.org/abs/2106.05452} {\path{arXiv:2106.05452}}.

\bibitem{helmig1997multiphase}
R.~Helmig, Multiphase flow and transport processes in the subsurface: a
  contribution to the modeling of hydrosystems., Springer-Verlag, 1997.

\end{thebibliography}
\bibliographystyle{elsarticle-num}
\newpage
\begin{appendices}

\section{Algorithm to compute the interface term integral}
\begin{algorithm}
\scriptsize
\caption{Computing integration points for a numerical source term integral over
a triangle $T \in \Gamma_h$. For each coupled network segment $\Lambda_i$, exactly one
integration point is computed. Accuracy is increased by local virtual refinement.}
\label{algo}
\begin{algorithmic}[1]
\Variables
\State \texttt{T}, a triangle with $3$ corner points, centroid $\vec{x}_T$ and area $A_T$
\State \texttt{lvlmax}, the maximum refinement level
\State \texttt{I}, an index triple of network segment indices
\State \texttt{Q}, an array of integration points (an integration point is a tuple $(\vec{x}, w, i)$
where $\vec{x} \in \mathbb{R}^3$ is a position, $w$ a weight, and $i$ the index of the coupled network segment)
\EndVariables
\Statex
\State \textbf{initialize:} \texttt{lvlmax} $\geq 0$, \texttt{T}, \texttt{Q} \Comment{\texttt{T} is on the coupling surface}
\State $\texttt{I} \gets \texttt{indices\_of\_closest\_segments}(\texttt{corners}(\texttt{T}))$
\State\Call{add\_integration\_points}{\texttt{T}, \texttt{lvlmax}, \texttt{I}, \texttt{Q}}
\For{$\texttt{q} \in \texttt{Q}$}
  \State $\texttt{q.position()} \gets \texttt{q.position()}/\texttt{q.weight()}$ \Comment{compute centroid}
\EndFor
\Statex
\Function{add\_integration\_points}{\texttt{T}, \texttt{lvl}, \texttt{I}, \texttt{Q}}
  \If{\texttt{all\_indices\_equal(I)}}
      \If{$ \exists \texttt{q} \in \texttt{Q} \,\text{such that}\, \texttt{q.index()} \in \texttt{I}$}
      \State $\texttt{q.position()} \gets \texttt{q.position()} + \vec{x}_T$
      \State $\texttt{q.weight()} \gets \texttt{q.weight()} + A_T$
      \Else
      \State $\texttt{Q.add\_new\_integration\_point}((\vec{x}_T, A_T, \texttt{I}[0]))$
      \EndIf
  \ElsIf{$\texttt{lvl} = \texttt{lvlmax}$}
      \For{$\texttt{corner} \in \texttt{T}$}
        \State $\texttt{idx} \gets \texttt{index\_of\_closest\_segment(\texttt{corner})}$
        \If{$\exists \texttt{q} \in \texttt{Q} \,\text{such that}\, \texttt{q.index()} = i$}
        \State $\texttt{q.position()} \gets \texttt{q.position()} + \texttt{corner}$
        \State $\texttt{q.weight()} \gets \texttt{q.weight()} + \frac{1}{3}A_T$
        \Else
        \State $\texttt{Q.add\_integration\_point}((\texttt{corner}, \frac{1}{3}A_T, \texttt{idx}))$
        \EndIf
      \EndFor
  \Else
    \For{$\texttt{TT} \in \texttt{refine(T)}$}
    \State $\texttt{II} \gets \texttt{indices\_of\_closest\_segments}(\texttt{corners}(\texttt{TT}))$
    \State\Call{add\_integration\_points}{\texttt{TT}, \texttt{lvl}+1, \texttt{II}, \texttt{Q}} \Comment{recursion}
    \EndFor
  \EndIf
\EndFunction
\end{algorithmic}
\end{algorithm}

\section{Analytical solutions for cylinder benchmark}
\label{app:benchmark}
Here, we give analytical expressions for the pressure solutions in the extended domain $\Omega_\text{ex}$
for the benchmark problem \cref{eq:kernel:problembeta}.
While the solution in the 1D domain $\Lambda$ and the exact source term $q$
is identical for all schemes, the pressure solutions $p_{\sthreed,\up{e}}^{\textsc{m}}$
slightly differs depending
on the chosen implicit interface method $M$, and are given by \citep{Koch2019a}
\begin{subequations}
  \label{eq:sol_implicit_methods}
  \begin{align}
  p_{\sthreed,\up{e}}^{\textsc{ls}} &=  - \frac{1 + \xThree}{2\pi} \ln r, \\
  p_{\sthreed,\up{e}}^{\textsc{css}} &= \begin{cases} - \frac{1 + \xThree}{2\pi} \ln {R} & r \leq {R}, \\ - \frac{1 + \xThree}{2\pi} \ln r & r > {R}, \end{cases} \\
  p_{\sthreed,\up{e}}^{\textsc{ds}} &= \begin{cases} - \frac{1 + \xThree}{2\pi} \left[ \frac{r^2}{2\varrho^2} + \ln\left(\frac{\varrho}{{R}}\right) - \frac{1}{2} \right] & r \leq \varrho, \\ - \frac{1 + \xThree}{2\pi} \ln r & r > \varrho, \end{cases}
  \end{align}
\end{subequations}
where $R$ denotes the tube radius, and $\varrho$ the distribution kernel radius for the uniform
cylindrical kernel function suggested in \citep{Koch2019a}.

\section{Parameter and vessel configuration for parallel vessel case}
\Cref{tab:tissue,tab:2d_multiple} provide the parameter values
and the vessel configuration for the numerical example in \cref{sec:2d-multiple}.
\begin{table}[H]
  \centering
  \caption{Parameter values and units for case \cref{sec:2d-multiple}.}
  \label{tab:tissue}
  \begin{tabular}{l l l}
    \toprule
    parameter & value & unit \\\midrule
    $\mu_B$ & \num{3e-3} & \si{\pascal\s} \\
    $\mu_I$ & \num{1e-3} & \si{\pascal\s} \\
    $K$ & \num{1e-17} & \si{\square\m} \\
    $K_r$ & \num{1e-11} & \si{\m\per\pascal\per\s} \\
    $\Delta\pi$ & \num{2633} & \si{\pascal} \\
    \bottomrule
  \end{tabular}
\end{table}
\begin{table}[H]
  \centering
  \caption{Vessel configuration for case \cref{sec:2d-multiple}. The domain $\Omega^\text{ex}$
  is given by the square $[-100,100]\times[-100,100]\si{\micro\m}$.
  Positive source terms $q_i$ signify fluid exerting vessels,
  negative source terms signify fluid absorbing vessels.}
  \label{tab:2d_multiple}
  \begin{tabular}{l l l r r}
    \toprule
    $i$ & $\vec{x}_i$ ($\times \SI{50}{\micro\m}$) & $R_i$ ($\times \SI{50}{\micro\m}$) & $p_{\soned,e,i}$ (\si{\pascal}) & $q_i (\si{\milli\g\per\day\per\milli\m})$  \\\midrule
    $1$ & $(-0.5, 0.866)$ & $0.25$ & $-800$ & $-0.0539047$\\
    $2$ & $(0.5, 0.866)$ & $0.15$ & $+600$ & $0.0239149$\\
    $3$ & $(-1.0, 0.0)$ & $0.08$ & $+600$ & $0.0127681$\\
    $4$ & $(0.0, 0.0)$ & $0.1$ & $+400$ & $0.0106153$\\
    $5$ & $(1.0, 0.0)$ & $0.2$ & $-400$ & $-0.0218049$\\
    $6$ & $(-0.5, -0.866)$ & $0.1$ & $+50$ & $0.0010997$\\
    $7$ & $(0.5, -0.866)$ & $0.23$ & $-200$ & $-0.0127440$\\
    \bottomrule
  \end{tabular}
\end{table}

\end{appendices}

\end{document}